\documentclass[journal=jacsat,manuscript=article,layout=twocolumn]{achemso}
\usepackage[version=3]{mhchem} 
\makeatletter
\newcommand*{\addFileDependency}[1]{\typeout{(#1)}
\@addtofilelist{#1}
\IfFileExists{#1}{}{\typeout{No file #1.}}
}\makeatother

\usepackage{url}            
\usepackage{booktabs}      
\usepackage{amsfonts}       
\usepackage{nicefrac}       
\usepackage{microtype}      
\usepackage{xcolor}         
\usepackage{amsmath}
\usepackage{mathtools}
\usepackage{braket}
\usepackage{natbib}
\usepackage{graphicx}
\usepackage{cite}
\usepackage{ftnxtra}
\usepackage{fnpos} 
\usepackage{subcaption}
\usepackage{hyperref}
\usepackage{oplotsymbl}
\usepackage[normalem]{ulem}

\DeclareMathOperator{\EX}{\mathbb{E}}

\DeclareMathOperator{\vx}{\boldsymbol{x}}

\DeclareMathOperator{\Ne}{N_{e}}
\DeclareMathOperator{\rhom}{\rho_{{\cal M}}}
\DeclareMathOperator{\rhomstar}{\rho^*_{{\cal M}}}
\DeclareMathOperator{\rhozero}{\rho_{0}}

\DeclareMathOperator{\Tphi}{T_{\phi}}
\DeclareMathOperator{\gphi}{g_{\phi}}
\DeclareMathOperator{\rhophi}{\rho_{\phi}}

\DeclareMathOperator{\vz}{\mathbf{z}}
\DeclareMathOperator{\hrhom}{\hat{\rho}_{{\cal M}}}
\DeclareMathOperator{\rhozeropm}{\tilde{\rho}_{0}}
\DeclareMathOperator{\lambdazero}{\lambda_0}

\author{Alexandre de Camargo}
\affiliation[Mc]{Department of Chemistry and Chemical Biology, McMaster University, Hamilton, ON, Canada}

\author{Ricky T. Q. Chen}
\affiliation[meta]{FAIR at Meta, NY, USA}

\author{Rodrigo A.~Vargas-Hernández}
    \email{vargashr@mcmaster.ca}
\affiliation[Mc]{Department of Chemistry and Chemical Biology, McMaster University, Hamilton, ON, Canada}
\alsoaffiliation[BIMR]{Brockhouse Institute for Materials Research, McMaster University, Hamilton, ON, Canada}

\title{Leveraging Normalizing Flows for Orbital-Free Density Functional Theory}

\abbreviations{ML, OF-DFT, Normalizing Flows}

\begin{document}

\begin{abstract}
Orbital-free density functional theory (OF-DFT) for real-space systems has historically depended on Lagrange optimization techniques, primarily due to the inability of previously proposed electron density approaches to ensure the normalization constraint. This study illustrates how leveraging contemporary generative models, notably normalizing flows (NFs), can surmount this challenge. We develop a Lagrangian-free optimization framework by employing these machine learning models for the electron density. This diverse approach also integrates cutting-edge variational inference techniques and equivariant deep learning models, offering an innovative reformulation to the OF-DFT problem.
We demonstrate the versatility of our framework by simulating a one-dimensional diatomic system, LiH, and comprehensive simulations of hydrogen, lithium hydride, water, and four hydrocarbon molecules. The inherent flexibility of NFs facilitates initialization with promolecular densities, markedly enhancing the efficiency of the optimization process.
\end{abstract}\\

\noindent\emph{Introduction:} The density functional theory (DFT) framework has evolved into an indispensable tool in both computational materials science and chemistry, with the Kohn-Sham (KS) formalism being the de facto (or most commonly employed) form of DFT~\citep{kohn1965_self_DFT,becke50yrs:2014,ksdft:truhlar:2016,mardirossian2017thirty}.
The success of the KS formalism sparked a race to develop exchange-correlation (XC) energy functionals based on electronic spin densities~\citep{dft30yrs:hgordon,dft:review:2022,perdew1996generalized:pbe,stephens1994ab:b3lyp,beck1993density:b3lyp,heyd2003hybrid:HSE06,borlido2020exchange:XC}. Initially, physics-motivated functionals were the predominant framework until machine learning (ML) approaches emerged, marking a noteworthy shift in the landscape of quantum chemistry~\citep{xc-bo:vargas:jpca:2020,xc-ml:Burke:prl:2021,xc-hole:jcp:2021,xc-ml:review2023,kirkpatrick2021pushing:dm21,ma2022evolving:gas22}.

Orbital-free DFT (OF-DFT), rooted in the Hohenberg-Kohn theorems~\citep{hohenberg1964_inhomogeneous,gpaar1980_DFT}, is a promising alternative to KS-DFT given its lower computational scaling.
However, the imperative for relative accuracy in kinetic energy (KE) functionals, comparable to the total energy, remains a primary impediment~\citep{zhang2024overcoming:impediment,karasiev2012issues:issues_and_chalanges}. 
Research endeavors have extensively explored the parametrization of KE functionals~\citep{mazo2023variational,hodges1973quantum:tf_approx,brack1976extended:tf_approx}, surpassing the original Thomas-Fermi-Weizäcker-based formulation. Notable extensions involve non-local KE functionals based on linear response theory, such as the Wang-Teter~\citep{wang1992kinetic:tf_approx}, Perrot~\citep{perrot1994hydrogen}, Wang-Govind-Carter~\citep{wang1998orbital}, Huang-Carter~\citep{huang2010nonlocal:tf_approx}, Smargiassi-Madden~\citep{smargiassi1994orbital}, Foley-Madden ~\citep{foley1996further} and Mi-Genova-Pavanello~\citep{mi2018nonlocal} functionals, showcasing the capability of OF-DFT in simulating systems with a large number of atoms.

Similar to the development of XC functionals, the pursuit of highly accurate OF-DFT simulations has driven the development of KE functionals through ML algorithms. Predominant approaches employ kernel ridge regression~\citep{scikit-learn}, convolutional neural networks~\citep{o2015introduction}, and ResNets~\citep{he2016deep}. Notably, data used for training ML-based KE functionals are generated through KS-based simulations. However, a key limitation in data-driven functionals lies in the accuracy of functional derivatives, which, when poor, can result in highly inaccurate densities.
Despite significant recent progress in materials modeling within the OF-DFT framework, which now includes ML techniques, a consistent aspect for real-space simulations has been the parametrized form of the trial electron density. These traditional approaches have forced the OF-DFT framework to be a Lagrangian-based scheme.
In this work, we propose an alternative approach employing generative models, specifically normalizing flows, circumventing the normalization constraints that affect traditional methods in the OF-DFT real-space setup.
Our research is motivated by previous efforts that aim to develop unconstrained optimization methods for computational chemistry \citep{ref2:alharbi2017kinetic,ref2:fan1991optimization,ref2:saad2010numerical,ref2:schlegel2011geometry,ref2:thogersen2004trust}.\\

\noindent\emph{Methods:} In the  OF-DFT framework, the ground state energy ($E_{\text{gs}}$) and electron density ($\rhom$) are determined by minimizing the total energy functional ($E[\rhom]$),
{\footnotesize
\begin{equation}  \label{eqn:rho_class}  
\begin{split}
    E_{\text{gs}} &= \inf_{\rhom \in \Omega} E[\rhom(\vx)], \\
  \Omega &= \left \{\rhom:\rhom \in X, \int \rhom(\vx) \mathrm{d}\vx = \Ne \right \},
\end{split}
\end{equation}
}
where $X$ is the admissible class of physically realizable densities for $\rhom$, satisfying the normalization constraint on the total number of particles $\Ne$.
The OF-DFT framework's resemblance to variational inference in machine learning~\citep{variationalinference:review} lies in their shared objective of approximating/learning a density distribution through an optimization/minimization procedure.
All previously proposed methodologies belong to the category of density models known as ``energy-based models"~\citep{enerybasedml:lecun,energy-basedmodels:2003}. For instance, $\rhom=f^{2}_{\phi}(\vx)/\int f^{2}_{\phi}(\vx) \mathrm{d}\vx$ or $\rhom=e^{-f_\phi(\vx)}/\int e^{-f_\phi(\vx)} \mathrm{d}\vx$. Common approaches for $f_\phi$ include multi-grid~\citep{bu2023efficient:multi_grid} and wavelet frameworks~\citep{natarajan2011:wavelets}, as well as a linear combination of atomic Gaussian basis sets~\citep{chan2001thomas}; here $\boldsymbol{\phi}$ is referred as model's parameters.
Although these frameworks are robust, they require the inclusion of a Lagrange multiplier ($\mu$) in the minimization objective, 
{\footnotesize
\begin{equation}
\min_{\rhom} E[\rhom(\vx)] -\mu \left( \int \rhom(\vx) \mathrm{d}\vx - \Ne \right),
  \label{eqn:min_of_dft}
\end{equation}
}
where $\mu$ is referred to as the chemical potential and is associated with the normalization constraint on $\Ne$. Typically, conventional methods for solving for $\rhom$ in real space involve self-consistent procedures based on functional derivatives, resulting in the Euler–Lagrange equation  $\delta E[\rhom(\vx)]/\delta \rhom(\vx) - \mu = 0$~\citep{gpaar1980_DFT}.\\

In this work, we introduce an alternative parameterization of the electron density, 
\begin{eqnarray}
    \rhom(\vx) := \Ne \rhophi(\vx),
    \label{eqn:rhom}
\end{eqnarray}
where $\rhophi$ is a normalizing flow (NF) model and is also referred to as the \emph{shape factor}~\citep{gpaar1980_DFT,gparr1983shapefactor}. 
This NF-based model allows us to reframe the OF-DFT variational problem as a Lagrangian-free optimization problem for molecular densities in real space, as the normalization is guaranteed by $\rhophi$.

In machine learning, NFs are common methodologies used for data generation and density estimation.
These generative models transform a base (simple) density distribution $\rhozero$ into a target (complex) density distribution ($\rho_\phi$) by leveraging the change of variables formula,
\begin{equation}
    \rhophi(\mathbf{x}) = \rhozero(\mathbf{z})\left|\text{det}  \nabla_{\vz} \Tphi(\vz) \right|^{-1},
    \label{eqn:density_transform}
\end{equation}
where $\Tphi$ is a bijective transformation\footnote[1]{$\Tphi: \mathbb{R}^{D} \xrightarrow{} \mathbb{R}^{D}$ is called a diffeomorphism, and it must be bijective, differentiable, and invertible.}. 
Eq.~\eqref{eqn:density_transform} guarantees the preservation of volume in the density transformation, while also allowing the computation of the target density in a tractable manner, making NFs a promising candidate for parameterizing $\rhom$. Additionally, automatic differentiation tools will enable the computation of high-order gradients of $\rhom$, commonly required in density functionals.

The proposed framework is rooted in optimal transport and measure theory where $\rhophi$ is known as the \emph{push-forward} of $\rhozero$ by the function $\Tphi$, denoted by $\rhophi = \Tphi\star\rhozero$~\citep{kobyzev2020normalizing_review}. 
In the context of generative models, $\Tphi$ is learned by minimizing metrics that measure the difference between the data distribution and the generative model. 
Here, $\Tphi$ will be optimized/learned by minimizing total energy functional, Eqs.~\ref{eqn:rho_class}-\ref{eqn:min_of_dft}.

In NFs, a common approach to parametrize $\Tphi$ is through a composition of functions; $\Tphi = T_{K} \circ \cdots \circ T_{1}$~\citep{papamakarios2021normalizing,rezende2015FineteFlow,kobyzev2020normalizing_review}. 
These composable transformations can be considered as a flow discretized over time. 
Discrete-time NFs were originally adapted by Cranmer et al.~\citep{cranmer:qflows} for $L^2-$Norm functions, making them well-suited for simulating quantum systems.  
Subsequent research has embraced this framework, exploring its applications across diverse domains. For instance, excited vibrational states of molecules~\citep{excitedstates:qlows}, quantum Monte Carlo simulations~\citep{waveflow:qflows,pfau:qflows,qmc:qflows,abinitio:xie2021ab}, and more recently for KS-DFT~\citep{ksdft:qflows}.

An alternative formulation of Eq.~\ref{eqn:density_transform}, proposed by \citet{chen2018neuralODe} and referred to as continuous normalizing flows (CNF), is centered around the computation of the log density, the score function ($\nabla_{\vx} \log \rho(\vx)$), and $\Tphi$ through a joint ordinary differential equation,
{\footnotesize
\begin{equation}    \label{eqn:cnf_evol}
\partial_t \begin{bmatrix}
\mathbf{z}(t) \\
\log \rhophi(\mathbf{z}(t)) \\
\nabla \log \rho_\phi 
\end{bmatrix} =\begin{bmatrix} 
\gphi(\mathbf{z}(t),t) \\
-\nabla \cdot \gphi(\mathbf{z}(t),t)\\
-\nabla^2 \gphi - \left(\nabla \log \rhophi\right)^T \left(\nabla \gphi(\mathbf{z}(t), t) \right)
\end{bmatrix}
\end{equation}
}
where ``$\nabla \cdot$'' denotes the divergence operator~\citep{chen2023odescore}. $\nabla_{\vx} \rho(\vx)$ can be computed using the "log-derivative trick", express as $\nabla_{\vx} \log \rho(\vx) = \nabla_{\vx} \rho(\vx)/\rho(\vx)$.
Note that, unlike discrete-time normalizing flows, this continuous-time formulation allows for the simultaneous computation of the samples, the density, and the score function, making it convenient for evaluating density functionals.
The joint computation of $[\mathbf{z}(t),\log \rhophi(\mathbf{z}(t)),\nabla \log \rho_\phi ]$ is achieved by solving an augmented ODE (Eq.~\ref{eqn:cnf_evol}), composed of three terms one for each component.
For more details regarding normalizing flows, we encourage the reader to consult Refs.~\citep{kobyzev2020normalizing_review,papamakarios2021normalizing}, and Section S.1 in the Supporting Information (SI).\\

Commonly, the total energy functional is composed of the addition of individual functionals,\vspace{-.2cm}
\begin{equation}
    E[\rhom] = T[\rhom] + V_{\text{H}}[\rhom] +  V_{\text{e-N}}[\rhom]  + E_{\text{XC}}[\rhom], \label{eqn:energy_func}
\end{equation}
where $T$ is the KE functional, $V_{\text{H}}$ is the Hartree potential, $V_{\text{e-N}}$ is the electron-nuclei interaction potential, and $E_{\text{XC}}$ is the so-called exchange and correlation (XC) functional. 
For this work, the KE functional is the sum of the Thomas-Fermi (TF) and Weizs\"{a}cker (W) functionals, $T[\rhom] = T_{\text{TF}}[\rhom] + \lambdazero T_{\text{W}}[\rhom]$, where the phenomenological parameter $\lambdazero$ was set to 0.2~\citep{chan2001thomas}. Other KE functionals are compatible with the proposed framework as long as they are differentiable.
The analytic equations of all functionals used here are reported in the SI.

\begin{figure}
    \centering
    \includegraphics[scale=.45]{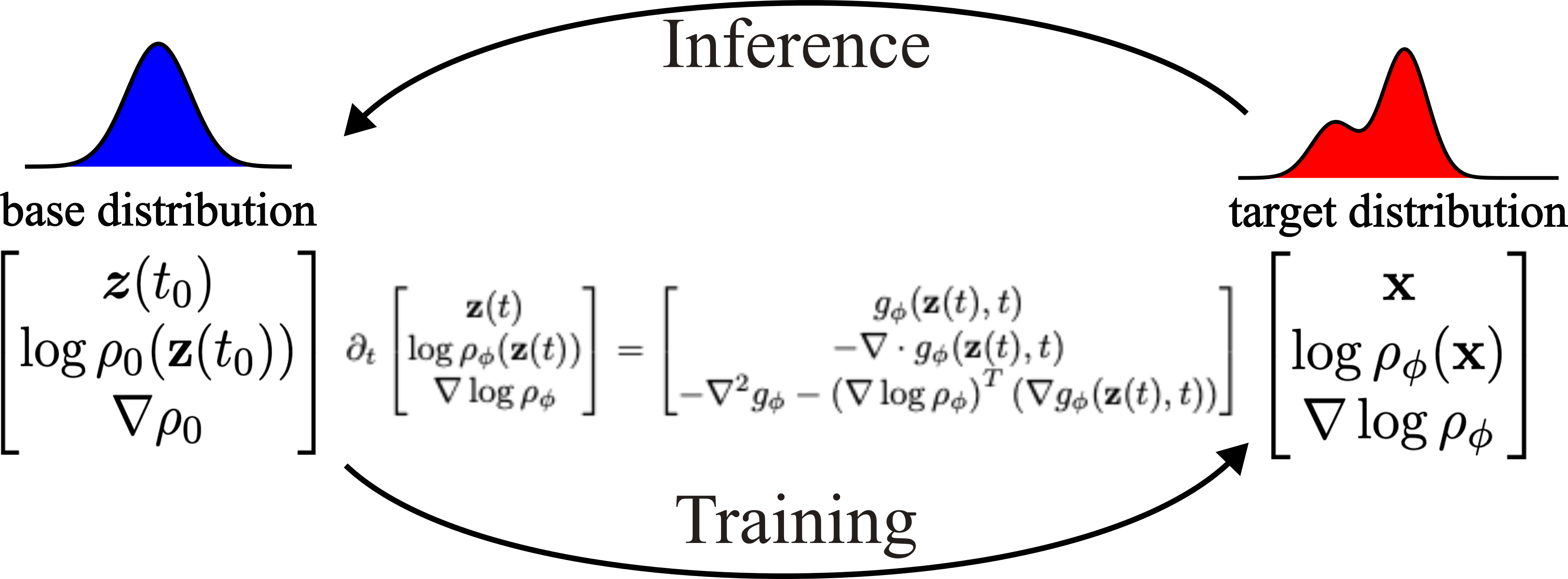}
    \caption{Continuous normalizing flows diagram.}
    \label{fig:cnf_diagram}
\end{figure}

For the proposed approach,  all individual density functionals are rewritten in terms of an expectation over the base distribution ($\rhozero$)~\citep{monte_carlo:mohamed2020monte,rezende2015FineteFlow},\vspace{-0.2cm}
{\footnotesize
\begin{equation}
\begin{split}
    F[\rhom] &= \int f(\vx,\rhom,\nabla\rhom)\rhom(\vx) \mathrm{d}\vx \nonumber \\
    &= (\Ne)^p \int f(\vx,\rhophi,\nabla\rhophi)\rhophi(\vx) \mathrm{d}\vx  \\
\end{split}
\end{equation}
}
{\footnotesize
\begin{equation}\label{eqn:expec_rho}
\begin{split}
    F[\rhom]&= (\Ne)^p \EX_{\rhozero} [ f(\Tphi(\vz), \rho_\phi,\nabla\rhophi) ],
\end{split}
\end{equation}
}
where $(\Ne)^p$ is the constant factor related to the number of electrons where $p\in\mathbb{R}^{+}$, and $f(\vx,\rho_\phi,\nabla\rho_\phi)$ is the integrand of the density functional $F[\rhom]$.
For all Monte Carlo (MC) estimates of the density functionals, the samples were drawn from the base distribution \footnote[2]{samples were drawn from $\rhozero$; $\vz \sim \rhozero$} and transformed by a CNF (Eq.~\ref{eqn:cnf_evol}), $\vx = \Tphi(\vz) := \vz + \int_{t_{0}}^{T} \gphi(\vz(t),t) \mathrm{d}t$; see Fig.~\ref{fig:cnf_diagram} and Section S.2 of the SI. We take a minibatch of samples and use Eq.~\ref{eqn:expec_rho} to construct an unbiased stochastic estimator of any density functional~\citep{rezende2015FineteFlow}. 

For the optimization of $\rhom$ (Eq.~\ref{eqn:min_of_dft}), one could use automatic differentiation to evaluate the gradient of the energy functional by differentiating through the numerical integration scheme. However, given the size of integration grids, this approach is resource-demanding and could lead to out-of-memory errors on many devices~\citep{ksdft:qflows}, particularly for the $V_{\text{H}}$ potential where a double integral is required. 

By reframing the OF-DFT problem as variational inference, we can use Eq.~\ref{eqn:expec_rho} to compute the expectation value of the energy functional and its gradient with respect to the model parameters\citep{rezende2015FineteFlow,monte_carlo:mohamed2020monte}, $\nabla_\phi E[\rhom] \approx \EX_{\rhozero} [\nabla_\phi f(\Tphi(\vz), \rhophi, \nabla\rhophi)]$, while guaranteeing the normalization constraint. Additionally, by integrating Eq.~\ref{eqn:cnf_evol} in the forward direction (noise-to-data), we can generate samples from the base distribution, and evaluate the density and its score function together. This approach supplies all the necessary components for evaluating a density functional, Fig.~\ref{fig:cnf_diagram}. 
Furthermore, our framework also allows using modern stochastic gradient optimization methods to minimize the total energy, homologous to existing variational inference algorithms~\citep{variationalinference:review,rezende2015FineteFlow}.
All required gradients were computed using the adjoint sensitivity method, as detailed in Ref.~\citep{chen2018neuralODe}, in \texttt{JAX}~\citep{jax2018github}, and the code developed for this work is available in the following \href{https://github.com/RodrigoAVargasHdz/ofdft_normalizing-flows}{repository}.

For a more accurate energy computation using CNFs standard quantum chemistry numerical integration techniques are beneficial. First, we integrate Eq.~\ref{eqn:cnf_evol} in the reverse direction (data-to-noise), mapping $\vx$ (grid point) to the base distribution space ($\vz$). This step is feasible because Eq.~\ref{eqn:cnf_evol} has a block structure, where only the first term is necessary to compute $\vz$ from $\vx$; $\vz = \Tphi^{-1}(\vx) := \vx + \int_{T}^{t_{0}} \gphi(\vz(t),t) \mathrm{d}t$. 
Given the known value of $\vz$, we compute the density and score function of the base distribution. Then, we integrate the equation forward in time, mirroring the training process, to evaluate the density functionals; Fig. \ref{fig:cnf_diagram}.
Combining these steps, we can compute the total energy, however, the complexity of these processes explains why directly differentiating numerical integration schemes for CNFs during training may not scale efficiently for larger systems.

In the context of our work, it is pertinent to note the application of automatic differentiation, a fundamental tool in the numerical ecosystem of deep learning libraries, and more recently in computational chemistry simulations~\citep{arrazola2023differentiable,arrazola2023HFQC,kasim2021dqmc,tamayo2018hfad,vargas2021adnes,vargashdz2023huxel,dawid2023modern,pyscfjax,schmidt2019machine:ad}. In OF-DFT simulations, noteworthy examples include \texttt{PROFESS-AD}~\citep{tan2023automatic}, and Ref.~\citep{DL:AD:OFDFT}, where functional derivatives, crucial for optimizing the electron density, were computed using \texttt{PyTorch}.\\

\begin{figure}[ht!]
    \centering
    \includegraphics[scale=0.27]{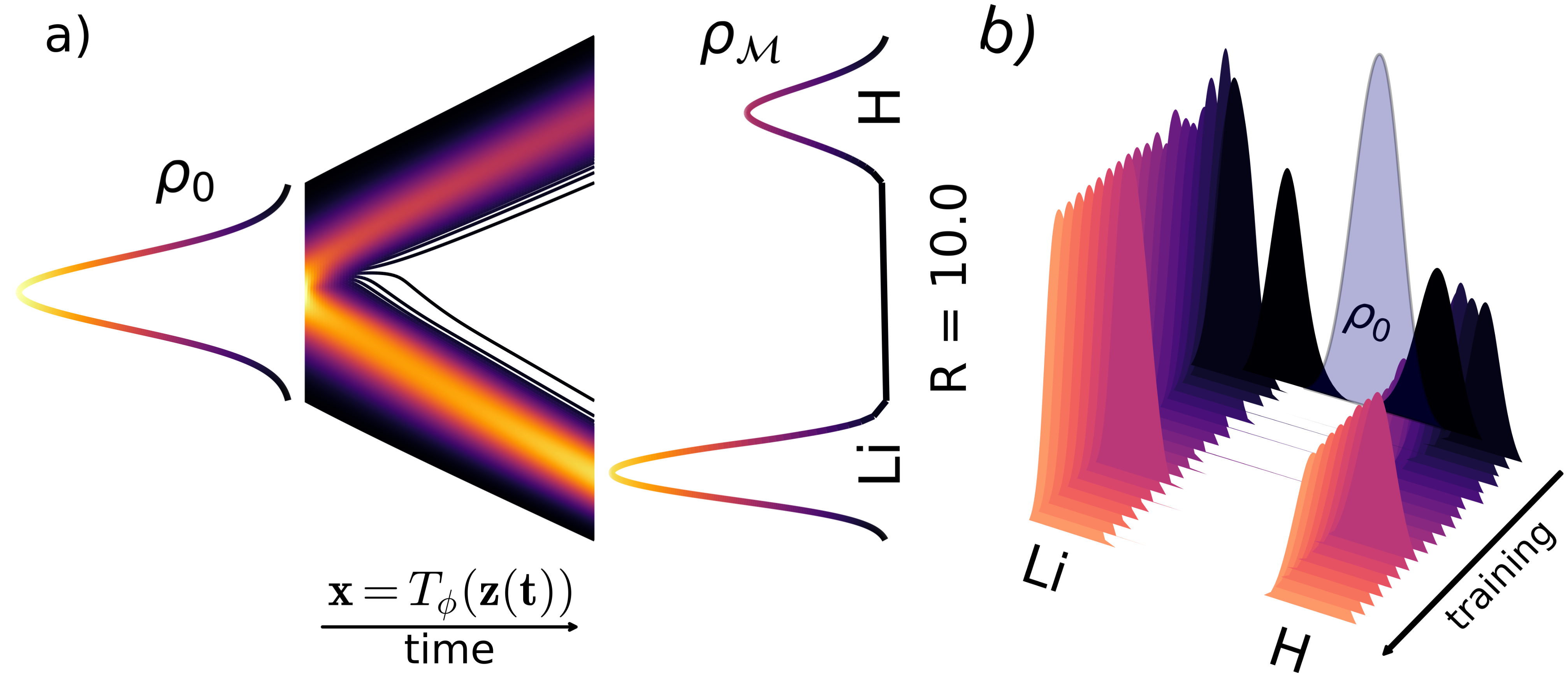}
    \caption{(a) The learned flow, Eq.~\ref{eqn:cnf_evol}, that minimizes the total energy for the LiH 1D system for $R=10$ a.u. (b) The change of $\rhom$ at different iterations of the optimization. $\rhozero$ is a zero-centered Gaussian distribution. See the text for more details of the simulations.}
    \label{fig:LiH_1D}
\end{figure}

\emph{Results:} To illustrate the parametrization of $\rhom$ through a CNF, we first considered a one-dimensional (1D) model for diatomic molecules based on Ref.~\citep{snyder2013orbital}. For this toy system, we considered the XC functional from Ref.~\citep{shulenburger2009spin}, and $T_{\text{W}}$ was computed using the score function through Eq.~\ref{eqn:cnf_evol}, $T_{\text{W}}[\rhom] = \frac{\lambdazero}{8} \int  \left(\nabla \log \rhom(x) \right)^2  \rhom(x) \mathrm{d}x$. 
The Hartree ($V_{\text{H}}$), and the external potentials ($V_{\text{e-N}}$)  both are defined by their soft version,~\citep{snyder2013orbital}
{\scriptsize
\begin{eqnarray}
    V_{\text{H}}[\rhom] &=& \int \int \tfrac{ \rhom(x)\rhom(x')}{\sqrt{1 + |x - x'|^2}} \mathrm{d}x \mathrm{d}x', \label{eqn:hartree_soft} \\
    V_{\text{e-N}}[\rhom] &=& -\int  \biggl( \tfrac{Z_\alpha}{\sqrt{1 + | x - R/2 |^2}} + \tfrac{Z_\beta}{\sqrt{1 + | x + R/2 |^2}} \biggr) \rhom(x) \mathrm{d}x. \label{eqn:ext_pot_soft}  \nonumber \\
\end{eqnarray}}
We chose LiH as the 1D diatomic molecule given the asymmetry due to the mass difference between its atoms; $Z_{\alpha} = 3$, $Z_{\beta} = 1$. We first considered the inter-atomic distance ($R$) equal to $10$ Bohr.
For the estimation of the total energy, we used 512 samples from the base distribution $\rhozero$, a zero-centered Gaussian distribution with $\sigma=1$. 
Fig.~\ref{fig:LiH_1D} illustrates the learned flow, or mass transport, from $\rhozero$ to $\rhom$ by the CNF (Eq.~\ref{eqn:cnf_evol}) that minimizes $E[\rhom]$, $\Ne = 2$. 
As we can also observe from Figs. \ref{fig:LiH_1D}-\ref{fig:pes_LiH_1D}, this CNF approach is capable of splitting the density given the large value of $R$ and allocating a higher concentration of electron density closer to the Li nuclei. 
Our simulations indicate that only $\sim 5,000$ optimization steps were needed for converged results, see Fig. S.1 in the SI.

\begin{figure}[!ht]
    \includegraphics[scale=0.3]{Figures/Final_LiH_1D/Fig3a-bLiH1D.png}
    \caption{(a) $\rhom$ for LiH for different nuclear distances $R$. For all simulations, $\rhozero$ is a zero-centered Gaussian distribution with $\sigma=1$.
    (b) The total energy of 1D LiH as a function of $R$. $V_{\text{NN}}(R)$ is the nuclear repulsion term and $R_e = 2.95 $ a.u. {\color{orange} $\pentagofill$}-symbols represent the total energy value computed with trapezoidal rule, and {\color{blue} $\bullet$}-symbols with MC.
    See the text for more details of the simulations.}
    \label{fig:pes_LiH_1D}
\end{figure}

We also explored the flexibility of our proposed CNF framework by examining various inter-nuclear distances for LiH, as shown in Fig.~\ref{fig:pes_LiH_1D}. For these 1D simulations, we consistently employed the same $\rhozero$, a 1D Gaussian distribution centered at zero. The neural network (NN) architecture used for $\gphi$ featured three hidden layers, each with 512 neurons and the $\tanh$ activation function. Other architectures were tested but were found to be sub-optimal.
Our simulations also reveal that $\gphi$, when randomly initialized, effectively accelerates the minimization of $E[\rhom]$, particularly at large inter-nuclear distances ($R \gg R_e$), where $R_e = 2.95$ Bohr denotes the equilibrium bond distance. These results demonstrate the flexibility of $\gphi$ for different scenarios, from strong nuclear interactions ($R<R_e$) to bond-breaking regimes ($R \gg R_e$), Fig.~\ref{fig:pes_LiH_1D}. 
The potential energy surface curve for the LiH, Fig.~\ref{fig:pes_LiH_1D}, further corroborates these findings.
We verified the validity of the proposed method by computing the total energy with the learned $\rhom$ using quadrature integration, finding no discernible difference between Adam and RMSProp except when $R\geq 6.0$ a.u. (Fig. S.1 and Table S.4 in the SI); however, in these 1D cases, RMSProp consistently achieves lower energy values (Table S.4 in the SI).\\

In normalizing flows, the transformation map $\Tphi$ (Eq.~\ref{eqn:density_transform}) connects the base density, $\rhozero$, with the target density, $\rhophi$. While $\rhozero$ is commonly modeled as a multi-variate Gaussian for applications like image generation, in the realm of molecular systems, adopting a promolecular density ($\rhozeropm$), emerges as a more natural base distribution. This choice enhances the base model's alignment with molecular structures and could potentially reduce the need for larger $\gphi$ models.
Here, we define $\rhozeropm = \sum_i c_i {\cal N}_i(\mathbf{R}_i,\sigma=1)$, where ${\cal N}_i$ is a \texttt{1S} orbital centered at the nucleus position ($\mathbf{R}_i$). The coefficients $c_i$ represent the proportional influence of each nucleus on the overall density, $\sum_i c_i = 1$, and $c_i = \tfrac{Z_i}{\sum_j Z_j}$ where $Z_i$ is the atomic number of the $i^{th}$-nucleus. 
Other base distributions could be considered given the computation of the density and the score function is tractable and samples can be easily generated.

To precisely model this density transformation and account for symmetries in the system, $\gphi$ is a permutation equivariant graph neural network (GNN)~\citep{equivflows:now:2020,perminvflows:wood:2023},
{\footnotesize
\begin{equation}
g_\phi(\mathbf{z},t) = \sum_i^{N_a} f_{\phi}(\|\mathbf{z}(t) - \mathbf{R}_i\|_2,\tilde{Z}_i)(\mathbf{z}(t)- \mathbf{R}_i),\label{eqn:equivnn}
\end{equation}}
where $N_a$ is the number of nuclei, $\tilde{Z}_i$ is the atomic number of the $i^{th}$-nucleus encoded as a one-hot vector, and $f_{\phi}$ is a $M$-layer NN with $64$ neurons per layer, and the $\tanh$ activation function. This GNN architecture is selected for its capability to process permutations of input atoms invariantly, thereby capturing the molecule's essential spatial and chemical properties, uninfluenced by the nuclei's order.
For $\rhom$ to be permutation invariant with respect to the atoms, the vector field ($\gphi$) must be permutation equivariant, and $\rhozero$ can be factorized across atoms, meaning permutation invariant~\citep{papamakarios2021normalizing,perminvflows:wood:2023}.

We further investigate the scalability of CNFs through simulations in realistic real-space systems, focusing on the H$_2$, LiH, H$_2$O, and four hydrocarbon molecules\footnote[3]{Simulations done on an NVIDIA V100 GPU.}. For these molecular systems, the exchange component of $E_{\text{XC}}$ was modeled using a combination of the local density approximation and the B88 exchange functionals. For the correlation component $E_{\text{C}}$, we utilized both the PW92 ~\citep{perdew1992accurate} and the VWN~\citep{vosko1980accurate,shulenburger2009spin:vwn} correlation functionals. Detailed equations are presented in the SI.

\begin{figure}[ht!]
    \centering
    \includegraphics[scale=0.3]{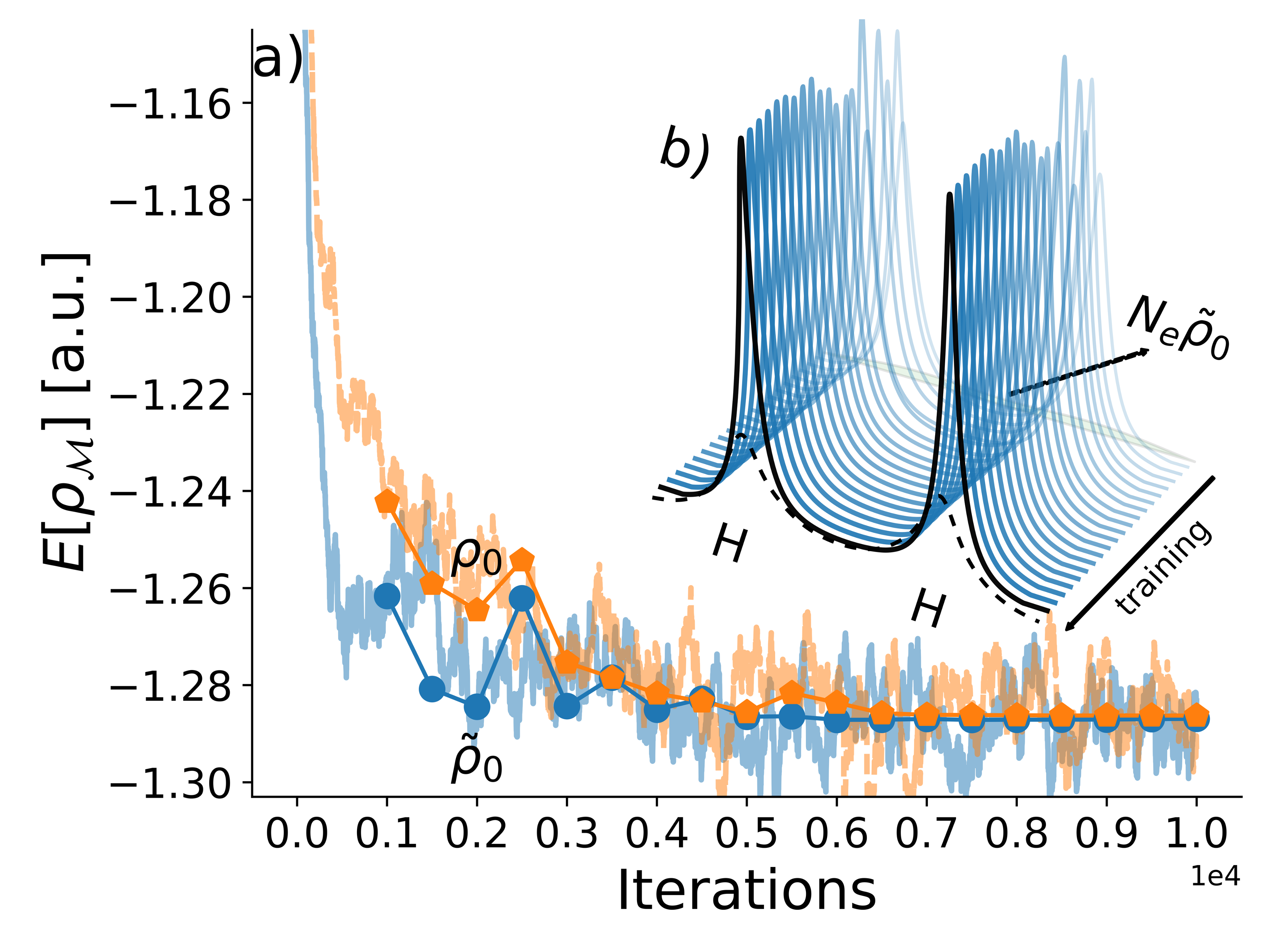}
    \includegraphics[scale=0.42]{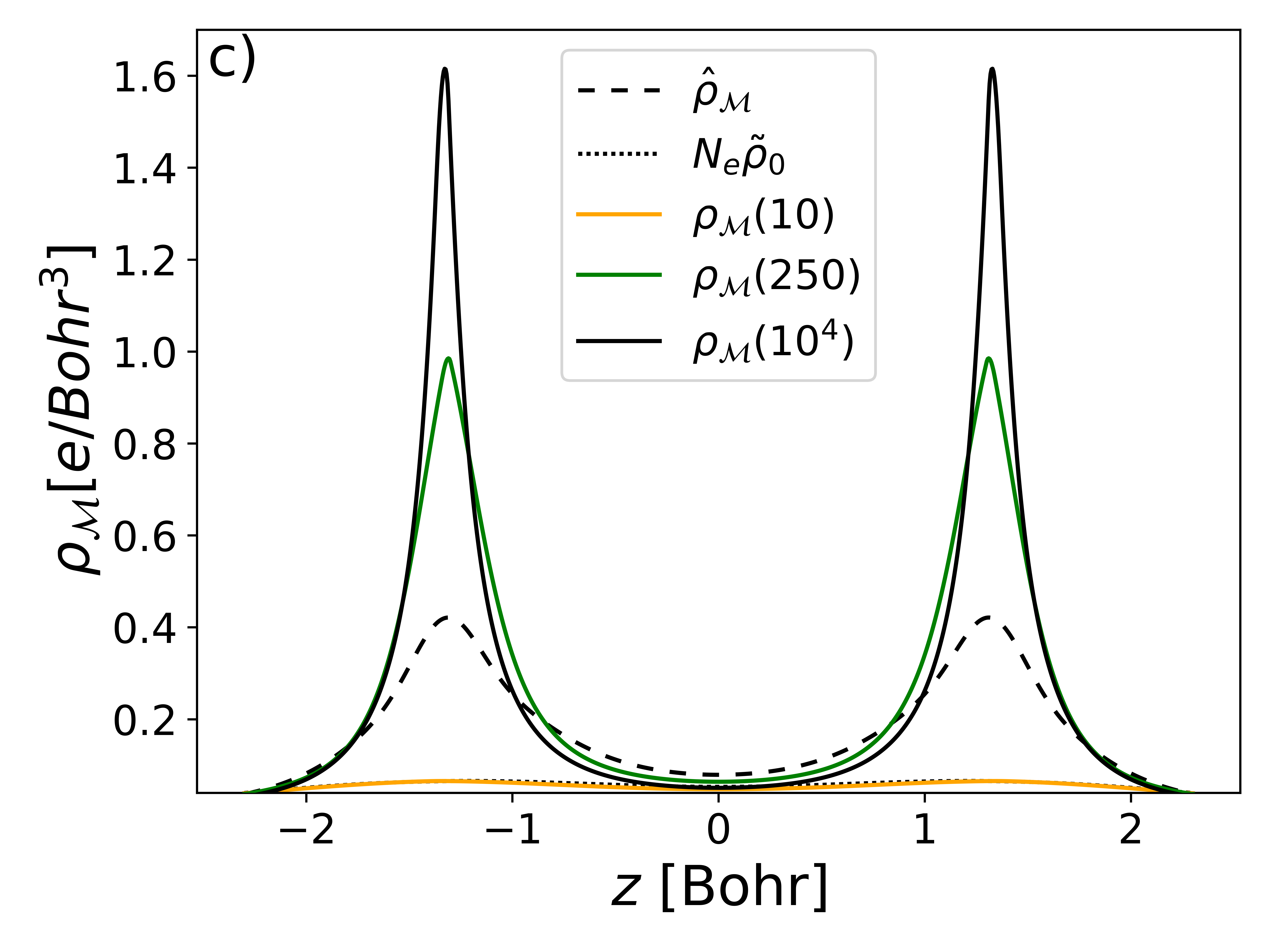}
    \caption{a) The total energy of H$_2$ molecule through the optimization for a CNF with a single Gaussian distribution ($\rhozero$) and a promolecular density ($\rhozeropm$). The symbols indicate the total energy computed with quadrature integration and the curves with Monte Carlo. 
    b) and c) The cross-section of the $\rhom$ at various iterations when $\rhozeropm$ is used. For these simulations, used the PW92 functional for the correlation functional. $\hrhom$ represents the density computed using the KS formalism with a 6-31G(d,p) basis set.
    }
    \label{fig:h2_3d_energy}
\end{figure}

For H$_2$ with $R = 0.7$ \AA, we found that $\sim 5,000$ iterations are needed for the total energy to stabilize, see Fig.~\ref{fig:h2_3d_energy}. 
We further validate the total energy value using quadrature integration (MC), -1.2835 (-1.2798) a.u. for the VWN functional, and -1.2837 (-1.2799) a.u. for the PW92 functional. 
The difference between utilizing $\rhozero$ or $\rhozeropm$ in this diatomic system is minor, Fig.~\ref{fig:h2_3d_energy}a. 
We also found a $\sim 1$ kcal/mol energy difference when $\gphi$ with an additional layer is considered; see Table S.6 in the SI. 
Additionally, Figs.~\ref{fig:h2_3d_energy}b and \ref{fig:h2_3d_energy}c illustrate the change of $\rhom$ through the optimization, notably showcasing an increase in the electron density around the nuclei. 
As a reference, the total energy for a KS-DFT simulation for the VWN functional with the 6-31G(d,p) (STO-3G) basis set is -1.6133 (-1.5917) a.u.
The results for LiH are presented in Table S.6 in the SI. 

For H$_2$O, the total energy stabilizes at $\sim 8,000$ iterations when using $\rhozeropm$. 
As opposed to H$_2$, we found a significant improvement for water when a three-layer GNN was used without a big compromise in the optimization time (see Table S.6 in the SI).
The total energy, computed with quadrature integration, for the VWN (PW92) functional, is -82.44516 (-82.48210) a.u.  
The results with $\rhozeropm$ and the proposed $\gphi$ architecture (Eq.~\ref{eqn:equivnn}) agree with a KS-DFT simulation using a minimal basis set, which yielded -83.9016 a.u. This energy discrepancy is expected given the level of the KE functional used in the simulations.
Additional information on the simulations is presented in the SI.

\begin{figure}[ht!]
    \centering
    \includegraphics[scale=0.3]{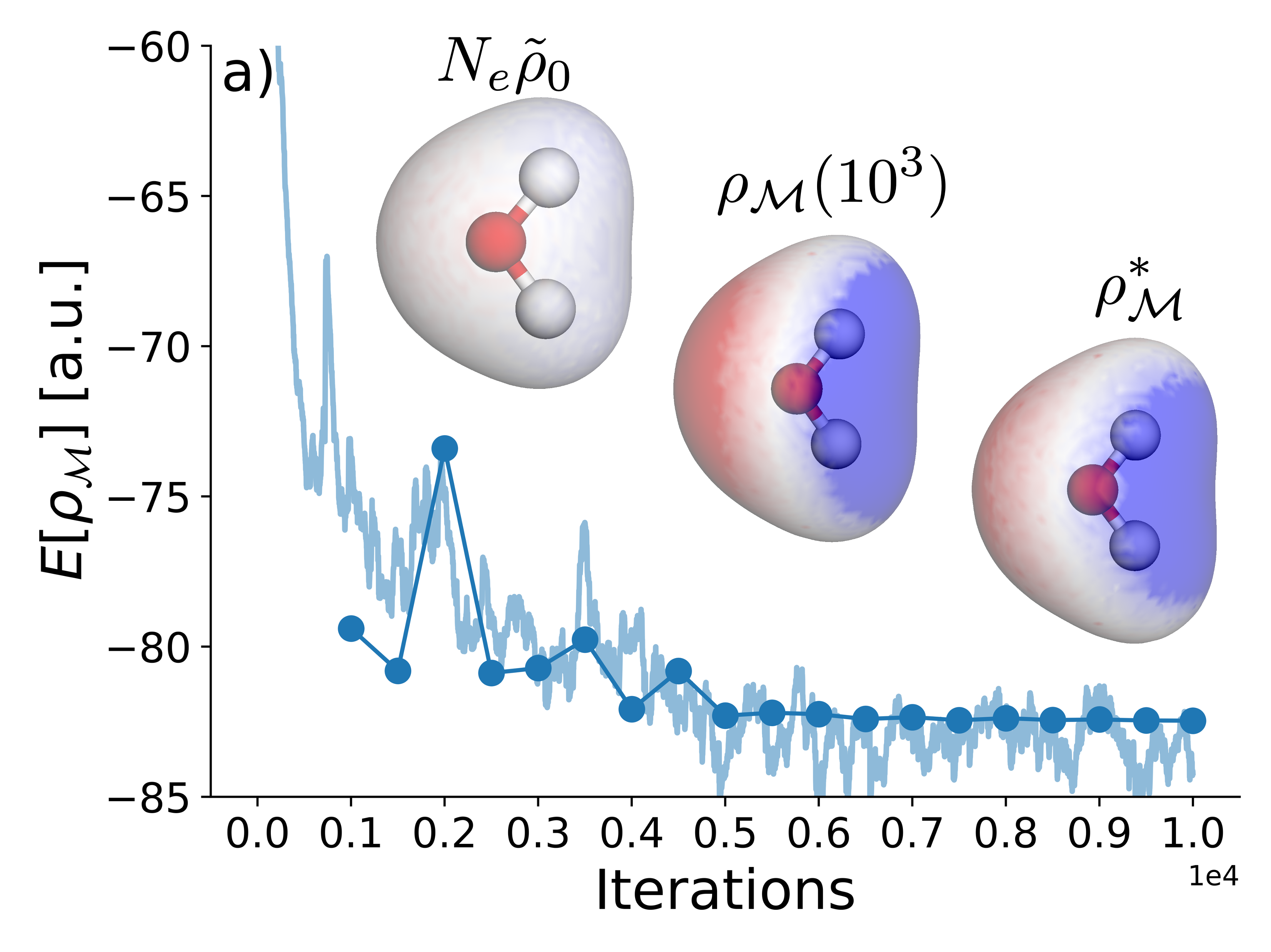}
    \includegraphics[scale=0.24]{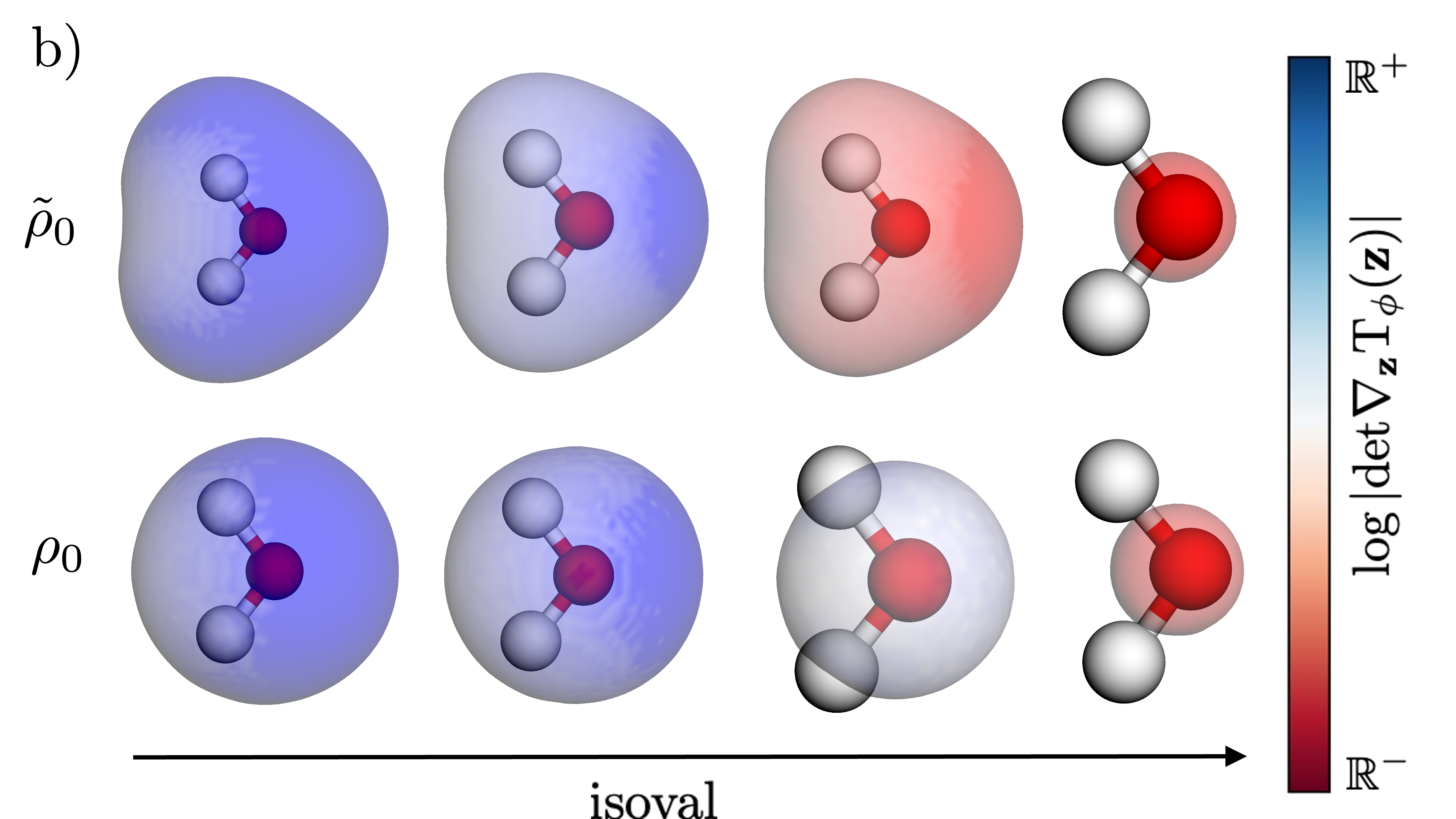}
    \caption{a) The total energy of H$_2$O molecule through the optimization for a CNF with a promolecular density ($\rhozeropm$) as the base distributions. The symbols indicate the total energy computed with quadrature integration and the curves with Monte Carlo. For $\rhom$, the color-coded map indicates the value of the molecular electrostatic potential, and $\rhomstar$ is the density with the lowest energy.
    b) The learned change of density ($\log\left|\text{det}  \nabla_{\vz} \Tphi(\vz) \right|)$ by the CNF (Eq.~\ref{eqn:density_transform}) at different values of the base distribution; $\rhozero$ is a single Gaussian distribution.  
    For these simulations, we used the PW92 functional and a three-layer NN with 64 neurons per layer.}
    \label{fig:h2o_3d_energy}
\end{figure}

In normalizing flow-based models, the target density ($\rhophi$) is derived by effectively ``morphing'' the base distribution into the target one. As the complexity of the diffeomorphism increases, a larger network is needed to capture accurately the $\rhozero\to\rhophi$ transformation.
For the molecular systems studied in this work, as expected, $\Tphi$ (Eq.~\ref{eqn:cnf_evol}) learns to primarily increase the electron density closer to the nucleus region, even if $\rhozero$ has no previous knowledge of the location of the nucleus. 
This is illustrated in Fig.~\ref{fig:h2o_3d_energy}b, which displays $\log \left|\text{det}  \nabla_{\vz} \Tphi(\vz) \right|$ mapped over the base distribution for the water molecule. 
Our findings indicate that in regions proximal to the nucleus, $\Tphi$ effectively enhances electron density, as indicated by the sign of $\log \left|\text{det} \nabla_{\vz} \Tphi(\vz) \right|$. In contrast, $\Tphi$ reduces the value of $\rhozero$ in more distant areas, guaranteeing normalization. Fig.~\ref{fig:h2o_3d_energy}b further illustrates that $\Tphi$ is unique for the base distribution used. \\

\begin{figure}
    \centering
    \includegraphics[scale=0.15]{Figures/Final_3D_mol/TrajectoryC6H6Promol.png}
    \includegraphics[scale=0.16]{Figures/Final_3D_mol/TrajectoryC6H6Unimol.png}
    \caption{
    The trajectories learned by $\Tphi$ for Benzene using the promolecular density (panel a) and a single Gaussian distribution (panel b) as base distributions. Each curve represents the transformation of $\vz$, a random sample from the base distribution, to $\vx$. The color contour plot represents the cross-section of the electron density parametrized by the CNF. In both simulations, $\gphi$ is a 4-layer GNN (Eq.~\ref{eqn:equivnn}) with 64 neurons per layer and the $\tanh$ activation function. The XC functional used was the PW92 functional.
    }
    \label{fig:benzene_trajectory}
\end{figure}

To assess the scalability of CNFs, we also considered benzene (C$_{6}$H$_{6}$), anthracene (C$_{14}$H$_{10}$), pyrene (C$_{16}$H$_{10}$), and coronene (C$_{24}$H$_{12}$) molecules. For these systems, we found Adam to be a more robust optimizer compared to RMSProp, as the total energy computed using quadrature integration yields a lower value. The values of the total energies and training step times are reported in Table S.6 in the SI. We also observed that the total energy stabilizes at $\sim 12,000$ iterations for these hydrocarbon molecules, see Fig. S.6 in the SI.

For benzene, the training step time is at least three times faster when $\rhozero$ is a promolecular density. This can be observed in Fig.~\ref{fig:benzene_trajectory}, where $\Tphi$ locally updates $\vx$ around the nuclei, compared to the "long trajectory" required when $\rhozero$ is a single Gaussian distribution. We expect a similar trend for larger molecules given the complexity of distributing the density across more nuclei.
For C$_{14}$H${_10}$, C$_{16}$H$_{10}$, and C$_{24}$H$_{12}$, the training step time ranges from 8 to 13 seconds per iteration for 3 layers and from 9 to 17 seconds for 4 layers. This iteration time depends on the batch size and the stiffness of the ODE in the backward pass. Only for coronene, our results indicate that a 3-layer GNN yields better results than one with an additional layer; this is due to the nature of the MC procedure, which is noisier for larger systems. We believe this could be alleviated using a multi-device approach, allowing the use of a larger number of samples from $\rhozeropm$, and employing ODE regularization schemes~\citep{kelly2020easynode,kidger21a,xu2023normalizing}. These limitations could inspire the development of additional tools for the proposed framework.\\

\noindent\emph{Summary:} In this study, we introduce an innovative framework that utilizes generative models, particularly continuous normalizing flows, to parameterize electron densities in real space within molecular systems. This approach marks a significant shift away from traditional Lagrangian-based formulation within the OF-DFT framework. It distinguishes itself by ensuring direct normalization through the use of the change of variable formula and merges the strengths of variational inference with modern approaches in machine learning optimization and automatic differentiation. Our methodology was tested across various chemical systems and combined with promolecular densities. This initialization step introduces prior physical knowledge into the model.

Furthermore, the integration of generative models into OF-DFT, along with the use of equivariant GNN, complemented by recent advancements in kinetic energy functional development~\citep{ofdft:naturecs:2024,sun2023machine,seino2018semi,fujinami2020orbital,manzhos2020data,seino2019semi,meyer2020machine,mazo2023variational,benavides2024orbital:future_work}, holds a promising new avenue for the simulation of molecular systems. This different direction circumvents the limitations associated with grid-based methods, paving the way for alternative modeling of chemical systems within the OF-DFT framework in real space.\\

\begin{acknowledgement}
The authors thank A. Aldossary, J. Davidsson, R. Armiento, and C. Benavides-Riveros for fruitful discussions. 
This research was enabled in part by support provided by the Digital Research Alliance of Canada and NSERC Discovery Grant No. RGPIN-RGPIN-2024-06594.
\end{acknowledgement}

\bibliography{library}

\end{document}


\begin{abstract}
The purpose of this supplemental material is to provide more details about the proposed work in the main draft. Section \ref{sec:normflows} presents an introduction to normalizing flows, and Sections \ref{sec:MC} and \ref{sec:algorithm} describe the numerical details of the simulations and physical systems. 
In Section \ref{sec:results}, we present additional results from the ones presented in the main text.
\end{abstract}

\section{Normalizing Flows}\label{sec:normflows} 
The central goal of our work is to introduce an alternative parametrization for the electron density $\rhom$ using normalizing flows (NFs)\citep{kobyzev2020normalizing_review}, $\rhophi$. For convenience, we define $\rhom$ as, 
\begin{equation}
    \rhom(\vx) := \Ne \rhophi(\vx),
    \label{eqn:rhom}
\end{equation}
where $\rho_\phi$ is also known as \emph{shape factor} ~\citep{gpaar1980_DFT,gparr1983shapefactor}. Eq.~\ref{eqn:rhom} guarantees the normalization to the number of electrons $\Ne$, if $\rhophi$ normalizes to one; $\int \rhophi(\vx) \mathrm{d}\vx = 1$. As explained in the main text, we parametrize $\rhophi$ using a normalizing flow \citep{kobyzev2020normalizing_review,papamakarios2021normalizing}. 
Normalizing flows provides a general way of constructing complex probability distributions from simple ones, using the change of variable formula \citep{papamakarios2021normalizing}, 
\begin{equation}
    \rho_\phi(\mathbf{x}) = \rhozero(\mathbf{z})\left|\text{det}  \nabla_{\vz} T_\phi(\vz) \right|^{-1},
    \label{smeqn:density_transform}
\end{equation}
where $T_\phi(\cdot)$ is a bijective differentiable transformation and $\rhozero(\vz)$ is the base distribution. Eq. ~\ref{smeqn:density_transform}  guarantees volume preservation in the density transformation. 
One can parametrize $T_\phi(\cdot)$ through a series of composable functions~\citep{papamakarios2021normalizing,rezende2015FineteFlow,kobyzev2020normalizing_review}, 
\begin{equation}
    T_\phi(\vz) = (T_{K} \circ \cdots \circ T_{1})(\vz). 
    \label{eqn:transformation}
\end{equation}
These composable transformations can be seen as a flow discretized over time, and when multiple ``layers'' are considered it can parametrize further complex distributions.

\subsection{Continuous Normalizing flow}
An alternative formulation of (Eq. ~\ref{smeqn:density_transform}) is to construct a flow that operates in the continuous domain \citep{chen2018neuralODe,chen2023odescore}, assuming that the state transition is governed by an ordinary differential equation (ODE). This alternative NF framework is called \emph{continuous normalizing flow} (CNF), and it computes the $T_\phi$, the $\log$ density, and the score function by solving a joint ODE, 
\begin{equation}    \label{eqn:cnf_evol}
\partial_t \begin{bmatrix}
\mathbf{z}(t) \\
\log \rhophi(\mathbf{z}(t)) \\
\nabla \log \rho_\phi 
\end{bmatrix} =\begin{bmatrix} 
\gphi(\mathbf{z}(t),t) \\
-\nabla_{\vx} \cdot \gphi(\mathbf{z}(t),t)\\
-\nabla^2 \gphi - \left(\nabla \log \rhophi\right)^T \left(\nabla \gphi(\mathbf{z}(t), t) \right), 
\end{bmatrix}, 
\end{equation}
where ``$\nabla \cdot$'' denotes the divergence operator, $\nabla^2$ is  $\nabla \cdot \nabla$, and $\nabla_{\vx} \log \rho(\vx)$ represents the score function~\citep{chen2023odescore}, which is relevant for computing $\nabla_{\vx} \rho(\vx)$ using the "log-derivative trick",
\begin{equation}
    \nabla_{\vx} \log \rho(\vx) = \frac{\nabla_{\vx} \rho(\vx)}{\rho(\vx)}.
    \label{eqn:log_derivative_trick}
\end{equation}

We compute all required gradients by solving a second augmented ODE \citep{chen2018neuralODe}. This approach applies to all ODE solvers and it's called the adjoint sensitivity method \citep{pontryagin2018mathematical}. 
For all results presented in this work, we used the mixed 4th/5th order Runge-Kutta integration method\citep{shampine1986some} implemented in \texttt{jax.experimental.ode.odeint}. 

For the one-dimensional simulations, of the diatomic molecules, Section \ref{sec:results_1d}, the architecture of $\gphi$ is a standard feed-forward neural network (NN),
\begin{eqnarray}
    \gphi(\mathbf{z}(t),t) = (f_{M}\circ \cdots \circ f_1) (\mathbf{z}(t),t), \label{eqn:fnn}
\end{eqnarray}
where $f_\ell(\cdot)$ is a linear layer followed by an activation function, and $M$ is the number of layers. 
For the simulation in three dimensions, Section \ref{sec:results_3d}, $\gphi$ is parametrized by a permutation equivariant graph NN (GNN), \citep{equivflows:now:2020,perminvflows:wood:2023}
\begin{equation}
\gphi(\mathbf{z},t) = \sum_i^{N_a} f_{\ell}(\|\mathbf{z}(t) - \mathbf{R}_i\|_2,\tilde{Z}_i)(\mathbf{z}(t)- \mathbf{R}_i), \label{eqn:equiv_nn}
\end{equation}
where $\tilde{Z}_i$ is the atomic number of the $i^{th}$-nucleus, encoded as a one-hot vector ($[0,\cdots,1_{i},\cdots,0]$), $N_a$ is the number of nucleus in the molecule, and $f_{\ell}(\cdot)$ is a feed-forward NN with $64$ neurons per layer, also with the $\tanh$ activation function. This GNN architecture was chosen due to its ability to handle the nuclei symmetries.

\section{Expectation values of density functionals}\label{sec:MC}
In this section, we present the general framework used to compute the value of the total energy functional $E[\rhom]$ for the one-dimensional and three-dimensional simulations portrayed in the main text. 
For any of the examples considered in this work, the total energy functional $E[\rhom]$ is given by~\citep{hohenberg1964_inhomogeneous,gpaar1980_DFT}, 
\begin{equation}
    E[\rhom] = T[\rhom] + V_{\text{H}}[\rhom] +  V_{\text{e-N}}[\rhom]  + E_{\text{XC}}[\rhom],
    \label{eqn:energy_func}
\end{equation}
where we approximate the total kinetic energy ($T[\rhom]$) by the sum of the Thomas-Fermi ($T_{\text{TF}}[\rhom]$) and the Weizs\"{a}cker ($T_{\text{W}}[\rhom]$) functionals~\citep{gpaar1980_DFT}. 
$V_{\text{H}}[\rhom]$ is the Hartree potential that describes the classical electron-electron repulsion, $V_{\text{e-N}}[\rhom]$ is the external potential, and $E_{\text{XC}}[\rhom]$ is the exchange-correlation (XC) functional \citep{gpaar1980_DFT}. 

We use a Monte Carlo (MC) method to estimate the value of all individual functionals, through the following generalization,
\begin{eqnarray}
    F[\rhom] &=& \int f(\vx,\rhom,\nabla\rhom)\rhom(\vx) \mathrm{d}\vx = (\Ne)^p\int f(\vx,\rhophi,\nabla\rhophi)\rhophi(\vx)\mathrm{d}\vx,
    \label{eqn:expec_rho}
\end{eqnarray}
where $(\Ne)^p$, $p\in\mathbb{R}$, is the constant factor related to the number of electrons $\Ne$, due to our definition of $\rhom$ (Eq. \ref{eqn:rhom}), and $f(\vx,\rho_\phi,\nabla\rho_\phi)$ is the integrand of the functional $F[\rhom]$. The expectation value of $F[\rhom]$ is taken with respect to our base distribution $\rhozero$ \citep{rezende2015FineteFlow}, 
\begin{align}
     F[\rhom] &=  (\Ne)^p  \EX_{\rhozero} [ f(\Tphi(\vz), \rho_\phi,\nabla\rho_\phi) ] \approx (\Ne)^p \frac{1}{N} \sum^N_{i=1} \EX_{\rhozero} \left [ f(\Tphi(\vz_i),\rhophi,\nabla\rhophi) \right ], 
     \label{eqn:general_expectation}
\end{align}
where $N$ is the samples drawn from $\rhozero$ and transformed by the CNF, (Eq.~\ref{eqn:cnf_evol}); this is known as the reparameterization trick. In the following sections, we present the analytic expressions for each functional used in this work for the one-dimensional and three-dimensional cases.

\subsection{One-dimensional density functionals}\label{subsc:1d_functionals}
For the one-dimensional systems, the kinetic energy functional is the sum of the Thomas-Fermi ($T_{\text{TF}}[\rhom]$)\citep{snyder2013orbital} and the Weizs\"{a}cker ($T_{\text{W}}[\rhom]$)~\citep{gpaar1980_DFT} functionals, 
\begin{eqnarray}
    T_{\text{TF}}[\rhom] &=& \frac{\pi^2}{24} \int \left(\rhom(x) \right)^{3} \mathrm{d}x \label{eqn:thomas_fermi_1d} \\
    T_{\text{W}}[\rhom] &=& \frac{\lambdazero}{8} \int \frac{(\nabla \rhom(x))^2}{\rhom} \mathrm{d} x  = \frac{\lambdazero}{8} \int  \left(\nabla \log \rhom(x) \right)^2  \rhom(x) \mathrm{d} x, \label{eqn:1d_Weizsacker}    
\end{eqnarray}
where the phenomenological parameter $\lambdazero$ was set to 0.2~\citep{chan2001thomas}.
For the $V_{\text{H}}[\rhom]$ and $V_{\text{e-N}}[\rhom]$ functionals, we used the soft approximation~\citep{snyder2012finding}, 
\begin{align}
    V_{\text{H}}[\rhom] &= \int \int \frac{ \rhom(x)\rhom(x')}{\sqrt{1 + |x - x'|^2}} \mathrm{d}x \mathrm{d}x', \label{eqn:hartree_1d}  \\ 
     V_{\text{e-N}}[\rhom] &= \int v_{\text{e-N}}(x) \rhom(x) \mathrm{d}x \nonumber \\ &= -\int  \left  ( \frac{Z_\alpha}{\sqrt{1 + | x - R/2 |^2}} + \frac{Z_\beta}{\sqrt{1 + | x + R/2 |^2}} \right )\rhom(x) \mathrm{d}x,\label{eqn:ext_pot_1d}
\end{align}
where $Z_\alpha$ and $Z_\beta$ are the atomic numbers and $R$ is the distance between the two nuclei. The $E_{\text{XC}}[\rhom]$ functional's form is given by \citep{helbig2011density}, 
\begin{equation}
     E_{\text{XC}}[\rhom] = \int  \epsilon_{\text{XC}} \rhom(x) \mathrm{d}x,
\end{equation}
where $\epsilon_{\text{XC}}$ is \citep{shulenburger2009spin}, 
\begin{equation}
    \epsilon_{\text{XC}} (\rs,\zeta) = \frac{\azeta + \bzeta \rs + \czeta \rs^{2}}{1 + \dzeta \rs + \ezeta \rs^2 + \fzeta \rs^3} + \frac{\gzeta \rs \ln[{\rs + \alphazeta \rs^{\betazeta} }]}{1 + \hzeta \rs^2}. 
    \label{eqn:eqn_exc_1d}
\end{equation}
For all one-dimensional simulations, we used, $\rs = \tfrac{1}{2\rhom}$ (Wigner-Seitz radius\citep{shulenburger2009spin}) and $\zeta = 0$ (unpolarized density). All the parameters of $\epsilon_{\text{XC}}$ are defined in Table S.\ref{tab:exc_1d_parameters}.
The expectation values of $T_{\text{TF}}[\rhom]$, $T_{\text{W}}[\rhom]$, $V_{\text{H}}[\rhom]$, $V_{\text{e-N}}[\rhom]$ and $\Exc[\rhom]$, accordingly to Eq.~\ref{eqn:general_expectation}, are, 
\begin{align}
    T_{\text{TF}}[\rhom] &= \frac{\pi^2}{24} \Ne^3 \EX_{\rhozero} \left[ (\rhophi(x))^{2} \right]\label{eqn:tf_1d}, \\
     T_{\text{W}}[\rhom] &= \frac{\lambdazero}{8} \Ne  \EX_{\rhozero}\left[\left(\frac{\nabla \rho_\phi(x)}{\rho_\phi(x)}\right)^2\right] = \frac{\lambdazero}{8} \Ne\EX_{\rhozero} \left[\left(\nabla \log \rhophi(x) \right)^2\right]\label{eqn:tw_1d},  \\
    V_{\text{H}}[\rhom] &= \Ne^2 \EX_{\rhozero} \left [ \frac{1}{\sqrt{1 + |x - x'|^2}} \right],\label{eqn:soft_coul} \\
    V_{\text{e-N}}[\rhom] &=  \Ne \EX_{\rhozero} \left[  - \frac{Z_\alpha}{\sqrt{1 + | x + R/2 |^2}} - \frac{Z_\beta}{\sqrt{1 + | x + R/2 |^2}} \right]\label{eqn:nuclei_1d},\\
    \Exc[\rhom] &= \Ne \EX_{\rhozero} [\epsilon_{\text{XC}}(\rs,\zeta)]\label{smeqn:exchange_1d}.
\end{align}
\begin{table}
    \centering
    \begin{tabular}{|c|c|}
    \hline
         $\az$ & $-0.8862269$\\
         $\bz$ & $-2.1414101$\\
         $\cz$ & $0.4721355$\\
         $\dz$ & $2.81423$\\
         $\ez$ & $0.529891$\\
         $\fz$ & $0.458513$\\
         $\gz$ & $-0.202642$\\
         $\hz$ & $0.470876$\\
         $\alphaz$ & $0.104435$\\
         $\betaz$ & $4.11613$ \\
    \hline
    \end{tabular}
    \caption{Parameter values of $\epsilon_{\text{XC}}$ (Eq. \ref{eqn:eqn_exc_1d}) in a.u., obtained from Ref.~\citep{shulenburger2009spin}.}
    \label{tab:exc_1d_parameters}
\end{table}

\subsection{Three-dimensional density funcitonals}\label{subsc:3d_functionals}
In this section, we present the functionals used for the full three-dimensional simulations for the H$_2$, LiH, H$_2$O, C$_6$H$_6$, C$_{14}$H$_{10}$,C$_{16}$H$_{10}$ and C$_{24}$H$_{12}$ molecules.
The analytic forms of the used density functionals are,   
\begin{align}
     T_{\text{TF}}[\rhom] &= \frac{3}{10}(3\pi^2)^{\frac{2}{3}} \int \left(\rhom(\vx) \right)^{5/3} \mathrm{d} \vx, \label{eqn:thomas_fermi} \\ 
      T_{\text{W}}[\rhom] &= \frac{\lambda}{8} \int \frac{(\nabla \rhom(\vx))^2}{\rhom(\vx)} \mathrm{d} \vx  = \frac{\lambda}{8} \int  \left(\nabla \log \rhom(\vx) \right)^2  \rhom(\vx) \mathrm{d}\vx, \label{eqn:weizsacker} \\ 
      V_{\text{H}}[\rhom] &= \frac{1}{2} \int \int \frac{ \rhom(\vx)\rhom(\vx')}{\sqrt{|\vx - \vx'|^2}} \mathrm{d}\vx \mathrm{d}\vx', \label{eqn:hartree} \\
        V_{\text{e-N}}[\rhom] &= \int v_{\text{e-N}}(\vx) \rhom(\vx) \mathrm{d}\vx = - \int   \left  ( \sum_i \frac{Z_i}{\|\vx - \mathbf{R}_i\|} \right )\rhom(\vx) \mathrm{d}\vx, \label{eqn:ext_pot}   
\end{align}
where $\mathbf{R}_i$ is the position of the $i^{\text{th}}-$nucleus. 
We took advantage of the "log-derivative trick" (Eq.~\ref{eqn:log_derivative_trick}) and defined $T_{\text{W}}[\rhom]$ (Eq.~\ref{eqn:weizsacker}) in terms on $\nabla \log \rhom$.
Furthermore, the expectation values of the above functions (Eqs.~\ref{eqn:thomas_fermi}--\ref{eqn:ext_pot}) are, 
\begin{eqnarray}
    T_{\text{TF}}[\rhom] &=& \frac{3}{10}(3\pi^2)^{\frac{2}{3}} \Ne^{2/3} \EX_{\rhozero} \left[  \rho_\phi(\vx) \right], \label{eqn:exp_TF}\\
    T_{\text{W}}[\rhom] &=& \frac{\lambda}{8} \Ne  \EX_{\rhozero}\left[\left(\frac{\nabla \rho_\phi(\vx)}{\rho_\phi(\vx)}\right)^2\right] = \frac{\lambda}{8} \Ne\EX_{\rhozero} \left[\left(\nabla \log \rhophi(\vx) \right)^2\right]\label{eqn:exp_TW}, \\
    V_{\text{H}}[\rhom] &=& \frac{\Ne^2}{2} \EX_{\rhozero} \left [ \frac{1}{\sqrt{|\vx - \vx'|^2}} \right]\label{eqn:exp_H}, \\
    V_{\text{e-N}}[\rhom] &=&  -\Ne \EX_{\rhozero} \left[ \sum_i \frac{Z_i}{\|\vx - \mathbf{R}_i\|} \right ]\label{eqn:exp_e-N}, 
\end{eqnarray}

The XC functional is composed of the sum of the exchange (X) and correlation (C) terms,
\begin{equation}
     E_{\text{XC}}[\rhom] = \int  \epsilon_{\text{XC}} \rhom(\vx) \mathrm{d}\vx = \int  \epsilon_{\text{X}} \rhom(\vx) \mathrm{d}\vx + \int  \epsilon_{\text{C}} \rhom(\vx) \mathrm{d}\vx.
\end{equation}

We report all different $\epsilonx$ and $\epsilonc$ used in the simulations,
\begin{align}
    \epsilon_{\text{X}}^{\text{LDA}} &= -\frac{3}{4} \left(\frac{3}{\pi}\right)^{1/3} \rhom(\vx)^{1/3}
    \label{eqn:x_Dirac} \\ 
    \epsilon_{\text{X}}^{\text{B88}}&= -\beta \frac{X^2}{\left(1 + 6 \beta X \sinh^{-1}(X) \right)} \rhom(\vx)^{1/3}, \label{eqn:x_b88}
\end{align}
\begin{align}
     \epsilon_{\text{C}}^{\text{VWN}} &= \frac{A}{2} \left\{ \ln\left(\frac{y^2}{Y(y)}\right) + \frac{2b}{Q} \tan^{-1} \left(\frac{Q}{2y + b}\right) +\nonumber \right.\\&- \left.\frac{b y_0}{Y(y_0)} \left[\ln\left(\frac{(y-y_0)^2}{Y(y)}\right) + \frac{2(b+2y_0)}{Q}\tan^{-1}  \left(\frac{Q}{2y+b}\right) \right] \right\}  \label{eqn:correlation_vwn} \\ 
     \epsilon_{\text{C}}^{\text{PW92}} &= -2A(1+\alpha_1 \rs) \ln \left[{1 + \frac{1}{2A(\beta_1 \rs^{1/2} + \beta_2 \rs + \beta_3 \rs^{3/2} + \beta_4 \rs^{2}}}\right]  \label{eqn:correlation_pw92}\
\end{align}
where $\rs = \left ( \frac{3}{4\pi \rhom}  \right)^{\frac{1}{3}}$ from Ref.\citep{perdew1992accurate}. 
For $\epsilon_{\text{C}}^{\text{VWN}}$, $y = \rs^{1/2}$, $Y(y) = y^2 + by + c$, $Q = \sqrt{4c - b^2}$, and the constants $b$, $c$ and $y_0$ are given in the Table S.\ref{tab:correlation_VWN}. 
The PW92's parameters are reported in Table S.\ref{tab:correlation_PW92}.
For $\epsilon_{\text{X}}^{\text{B88}}$, $\beta$ is $0.0042$ a.u. and $X = \tfrac{|\nabla \rhom |}{\rhom^{4/3}}$~\citep{becke1988density}, where we use the  "log-derivative trick" (Eq. \ref{eqn:log_derivative_trick}) and expand in terms of the score function $X = \tfrac{|\rhom \nabla \log{\rhom} |}{\rhom^{4/3}}$. 

\begin{table}
    \centering
    \begin{tabular}{|c|c|}
    \hline
         $A$ & $0.0621814$ \\
         $y_0$  & $-0.10498$\\
         $b$ & $3.72744$\\
         $c$ & $12.9352$\\
    \hline
    \end{tabular}
    \caption{Parameter values of $\epsilon_{\text{C}}^{\text{VWN}}$ (Eq.~\ref{eqn:correlation_vwn}) in a.u., obtained from Ref.\citep{vosko1980accurate}.}
    \label{tab:correlation_VWN}
\end{table}

\begin{table}
    \centering
    \begin{tabular}{|c|c|}
    \hline
         $A$ & $0.031091$\\
         $\alpha_1$ & $0.21370$\\
         $\beta_1$ & $7.5957$\\
         $\beta_2$ & $3.5876$\\
         $\beta_3$ & $1.6382$\\
         $\beta_4$ & $0.49294$\\
    \hline
    \end{tabular}
    \caption{Parameter values of $\epsilon_{\text{C}}^{\text{PW92}}$ (Eq.~\ref{eqn:correlation_pw92}) in a.u., obtained from Ref.\citep{perdew1992accurate}.}
    \label{tab:correlation_PW92}
\end{table}

\section{Optimization algorithm}\label{sec:algorithm}
In this section, we present the optimization algorithm presented in the main text. 

The optimization of $\rhom$ was performed through an MC scheme where the parameters of the normalizing flow ($\phi$) are updated via a stochastic gradient optimization,
\begin{equation}
    \phi = \phi - \alpha \nabla_\phi E[\rhom], 
    \label{eqn:sgd}
\end{equation}
where $\alpha$ is the learning rate, and $\nabla_{\phi} E[\rhom]$ is the gradient of the energy with respect to the parameters $\phi$. As mentioned in the main text, $\nabla_{\phi} E[\rhom]$ is also estimated through a MC scheme \citep{monte_carlo:mohamed2020monte}, 
\begin{equation}
    \nabla_\phi E[\rhom] = \EX_{\rhozero} [\nabla_\phi f_{E}(\Tphi(\vz), \rhophi, \nabla\rhophi)] \approx \tfrac{1}{N}\sum_i^N \nabla_\phi f_{E}(\Tphi(\vz_i), \rhophi(\vz_i), \nabla_{\vz_{i}}\rhophi). 
    \label{eqn:gradient_energy}
\end{equation}
For this work, all required gradients were computed using automatic differentiation, as it is a common practice in computational chemistry~\citep{arrazola2023differentiable,kasim2021dqmc,tamayo2018hfad,vargashdz2023huxel}.

The proposed algorithm is depicted in Algorithm \ref{alg:cap}. 
For all simulations, we employed RMSProp~\citep{hinton2012rmsprop} and Adam~\citep{kingma2014adam} optimizers, featuring a learning rate schedule with initial and final values set at $3\times 10^{-4}$ and $10^{-7}$, respectively. 
Code was developed using \texttt{JAX} ecosystem~\citep{jax2018github,flax2020github,deepmind2020jax}. The code developed for this work is available in the following \href{https://github.com/RodrigoAVargasHdz/ofdft_normalizing-flows}{repository.}

\begin{algorithm}
\caption{Optimization Algorithm}\label{alg:cap}

    \begin{algorithmic}
        \Require CNF parameters $\phi$, base distribution $\rhozero$, energy functional $E[\cdot]$  
        \While{not converged}

            {\footnotesize
            $\{\vz_i, \log \rhozero(\vz_{i}), \nabla_z \log \rhozero(\vz_{i})\}_i^{N}$ $\sim \rhozero(\vz)$  \Comment{Sample the base distribution}
    
            {\footnotesize $[\vx_{i},\log \rhophi(\vx_{i}),\nabla_{\vx} \log \rhophi(\vx_{i})] = \texttt{ODESolve}([\vz_i, \log \rhozero(\vz_{i}), \nabla_z \log \rhozero(\vz_{i})],t_0,t_1,\phi$)}\Comment{Solve ODE Eq.(~\ref{eqn:cnf_evol})}
    
            $E =  \Ne^p\EX_{\rhophi}[E[\vx,\rhom,\nabla\rhom]] \approx \frac{\Ne^p}{N} \sum^N_{i} f_{E}(\vx_i,\rhophi(\vx_i),\nabla\rhophi(\vx_i))$ \Comment{Compute energy, Eq. (~\ref{eqn:expec_rho})}

            $\nabla_\phi E \approx \frac{(\Ne)^p}{N} \sum^N_{i} \nabla_\phi f_{E}(\vx_i,\rhophi(\vx_i),\nabla\rhophi(\vx_i))$ \Comment{Evaluate gradients, Eq.(~\ref{eqn:gradient_energy})}
    
            $\phi'$ = \texttt{optimizer step}($\phi$,  $\nabla_\phi E$) \Comment{Update parameters}}
            
        \EndWhile
    \State \Return parameters $\phi'$ 
    \end{algorithmic}
    \label{alg:opt_alg}
\end{algorithm}

\section{Results}\label{sec:results}
In this section, we present the results of the 1D and 3D simulations. 
The analytic equations of all functionals are presented in Sections \ref{subsc:1d_functionals} and \ref{subsc:3d_functionals}, and Section \ref{sec:algorithm} details the information of the optimization algorithm.  

\subsection{1D: LiH}\label{sec:results_1d}
In this section, we present additional results for the one-dimensional (1D) model of LiH at different inter-atomic distances ($R$). 
For all simulations, $\gphi$ (Eq. \ref{eqn:fnn}) was parametrized by a three-layer neural network with 512 neurons per layer, and a $\tanh$ function.
A maximum of 10,000 gradient iterations was allowed, and 512 samples from $\rhozero$.
We only considered as based distribution $\rhozero$ a 1D Gaussian distribution centered at $0$; ${\cal N}(0, \sigma=1)$. 

\begin{figure}[ht!]
    \centering
    \includegraphics[scale=0.7]{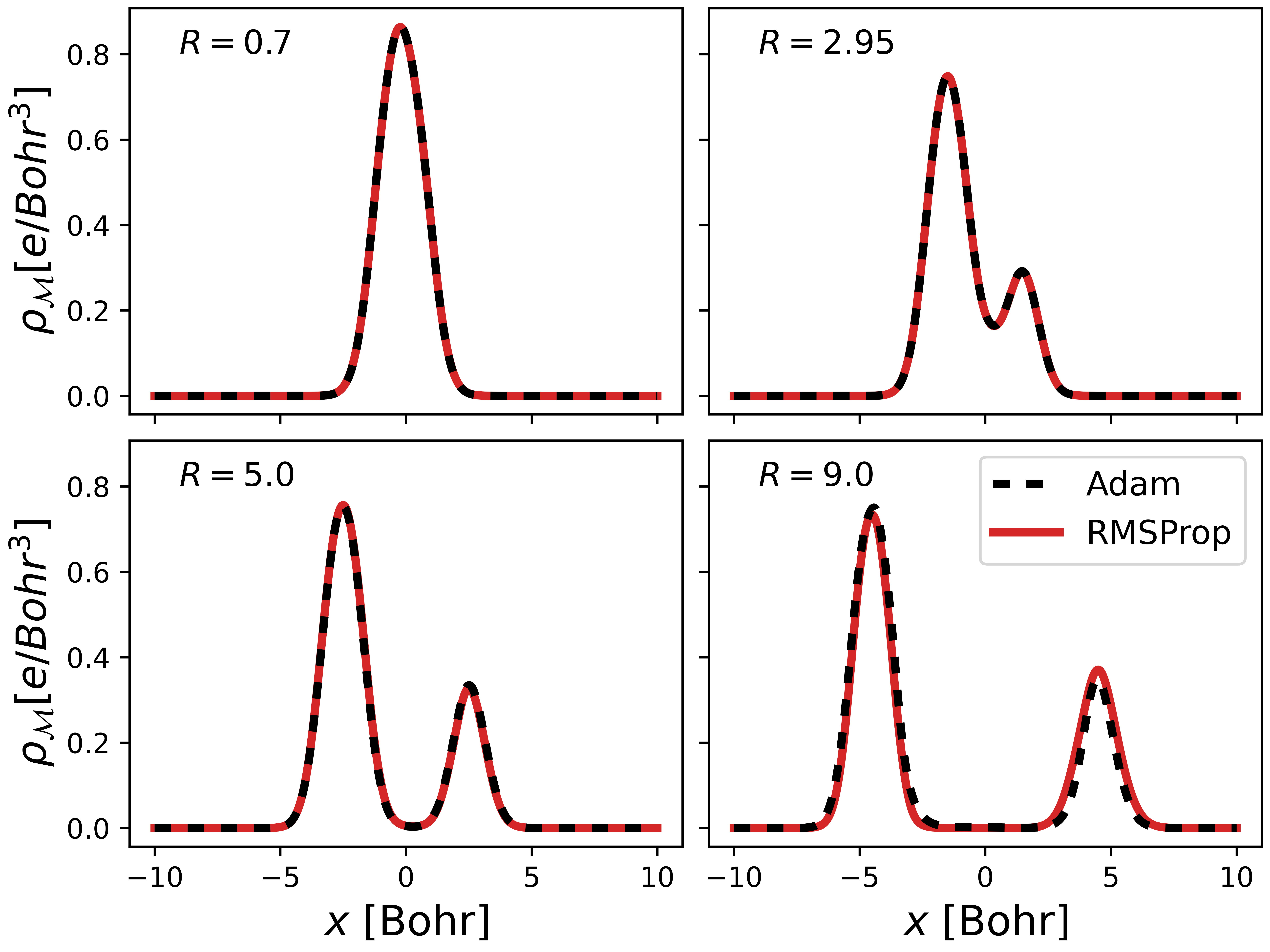}
    \caption{The plots of the electron density ($\rhom$), reported in electrons per cubic bohr, computed with RMSProp and Adam algorithms. Each panel shows the overlap between two independent simulations for different values of $R$. }
    \label{fig:mosaic_LiH1D}
\end{figure}

The energy functional for different values of $R$ was composed of the sum of $T_{\text{TF}}[\rhom]$, $T_{\text{W}}[\rhom]$, $V_{H}[\rhom]$ and $V_{\text{e-N}}[\rhom]$ (Eqs.~\ref{eqn:tf_1d}-\ref{eqn:nuclei_1d}). For the XC functional, we used the one from Ref. \citep{helbig2011density} (Eq.~\ref{eqn:eqn_exc_1d}).
In Fig. S.\ref{fig:ema_1D_LiH}, we presented the value of each functional through the optimization. To validate the total energy estimated with MC, we also used quadrature integration (trapezoidal rule). As observed, it follows the same trend as the estimated with MC.
For low values of $R$, the optimization converges rapidly as the $\rhom$ mimics a unimodal distribution. However, for larger values of $R$, a greater number of iterations is needed for $\gphi$ to split $\rhom$ into a bimodal distribution. Fig S.\ref{fig:mosaic_LiH1D} compares the $\rhom$ plots between RMSProp and Adam optimizers for various values of $R$.
Finally, the RMSProp algorithm consistently achieves lower energy values for this 1D system at different values of $R$, see Table S.\ref{tab:1d_energies}. 
For all 1D simulations, we used an NVIDIA Tesla P100 GPU.

\begin{figure}[ht!]
    \centering
    \includegraphics[scale=0.4]{Figures/SM/Ema1DMosaic.png}
 \caption{The value of the total energy and each functional, computed with MC (Eq.~\ref{eqn:general_expectation}), at each iteration. Each panel represents an independent simulation for different values of $R$. The {\color{orange} $\pentagofill$}-symbols represent the value of the total energy computed with quadrature integration. For all simulations, we used the same architecture for $\gphi$ and optimizer, see the text for more details. }
 \label{fig:ema_1D_LiH}
\end{figure}

\begin{table}
\begin{center}
\begin{tabular}{|c|c c |c c|c|}
   \hline
   & \multicolumn{2}{c|}{MC integration}&\multicolumn{2}{c|}{Quadrature integration}&\\
   \hline
   $R$ [a.u.]  & Adam & RMSProp & Adam & RMSProp &{$\Delta\epsilon  $ [kcal/mol]}\\
   \hline
   $0.7$ & $-4.16876$ & $ -4.16877$ & $-4.178977$&$-4.178976$ &$ 8.90 \times 10^{-4}$ \\
   \hline
    $0.8$ & $-4.14241$ & $-4.14243$ & $-4.152597$ &$-4.152596$&$4.62\times 10^{-4}$\\
   \hline
    $0.9$ & $-4.11384$& $-4.11386$& $-4.123992$ &$-4.123991$ &$6.50 \times 10^{-4} $\\
   \hline
    $1.0$ & $-4.08343$ &$-4.08345$ & $-4.093545$ &$-4.093542$ &$1.71\times 10^{-3} $\\
   \hline
    $1.1$ & $ -4.05158$ & $-4.05159$ & $-4.06164$ & $-4.06163$ &$2.14 \times 10^{-3} $ \\
   \hline
    $1.2$ & $-4.01864$ & $-4.01864$&$-4.02864$ & $-4.02863$ &$3.55 \times 10^{-3}$ \\
   \hline
    $1.4$ &$-3.95087$ &$-3.95088$ & $-3.960741$ & $-3.960740$ &$2.36 \times 10^{-3}$\\
   \hline
    $1.5$ & $-3.91668$ & $-3.91669$&$-3.9264736$ & $-3.9264734$ &$6.09 \times 10^{-5}$ \\
   \hline
    $2.0$ &$-3.75270$ & $-3.75271$&$-3.762070$& $-3.762071$ &$3.26 \times 10^{-3}$ \\
   \hline
    $2.5$ &$-3.61524$ &$ -3.61526$ &$-3.624130$ & $-3.624131$ &$2.58 \times 10^{-3}$ \\
   \hline
    $2.95$ &$-3.52041$ & $-3.52044$ &$-3.528893$& $-3.528896$ &$1.69 \times 10^{-3}$ \\
   \hline
    $3.5$ & $-3.43420$& $-3.43428$ &$-3.44234$ & $-3.44235$ & $5.32 \times 10^{-3}$\\
   \hline
    $4.0$ &$-3.37679$ & $-3.37686$ & $-3.3846610$ & $-3.3846618$ &$5.20 \times 10^{-4}$ \\
   \hline
    $5.0$  & $-3.30095$ & $-3.30151$&$-3.30903$ & $-3.30947$ & $0.276 $ \\
   \hline
    $6.0$ &$-3.25300$ &$-3.21186$ &
    $-3.26111$& $ -3.25537$ & $3.62$ \\
   \hline
    $7.0 $&$-3.14716$ &$-3.19113$ & $-3.1569$& $-3.18444$ &$17.4$ \\
   \hline
    $8.0$ &$-3.15803$ & $-3.15190$ &$-3.16708$&  $-3.16004$ &$4.42 $\\
   \hline
   $ 9.0$ &$-3.14766$ & $-3.12877$ & $-3.15686$& $-3.17489$ &$11.4$ \\
   \hline
    $10.0$ &$-3.11951$ & $ -3.12622$ & $-3.12898$ & $-3.13182$ &$1.78 $ \\
   \hline
\end{tabular}
\end{center}
\caption{The total energy, reported in Ha, is computed with MC and quadrature integration. $\Delta\epsilon$ is the absolute difference of the energy, computed using trapezoidal rule, between both optimization algorithms.} 
\label{tab:1d_energies}
\end{table}

\break
\subsection{3D:}\label{sec:results_3d}
In this section, we present additional results for the three-dimensional (3D) simulations for the H$_2$, LiH, H$_2$O, C$_6$H$_6$, C$_{14}$H$_{10}$, C$_{16}$H$_{10}$ and C$_{24}$H$_{12}$ molecules. 
Only for the small molecules, we considered two distinct base distributions, (i) a single Gaussian distribution ($\rhozero$), and (ii) a promolecular density ($\rhozeropm$),
\begin{eqnarray}
    \rhozeropm = \sum_i c_i {\cal N}_i(\mathbf{R}_i,\sigma=1),
    \label{eqn:promolec_density}
\end{eqnarray}
 where ${\cal N}_i(\cdot)$ is a \texttt{1S} orbital centered at the nucleus position, $\mathbf{R}_i$. The coefficients $c_i$ represent the proportional influence of each component on the overall density, $c_i  = \tfrac{Z_i}{\sum_j Z_j}$, and it depends on $Z_i$, the atomic number of the $i^{th}$-nucleus. For the hydrocarbon molecules (C$_{x}$H$_{y}$ molecules) we only consider a promolecular density.
To account for possible symmetries in the $V_{\text{e-N}}$ potential, $\gphi$ is a permutation equivariant graph neural network (GNN)\citep{equivflows:now:2020,perminvflows:wood:2023}, Eq.~\ref{eqn:equiv_nn}.
For all 3D simulations, we used an NVIDIA V100 GPU.

\begin{figure}[!ht]
    \centering
    \includegraphics[scale=0.3]{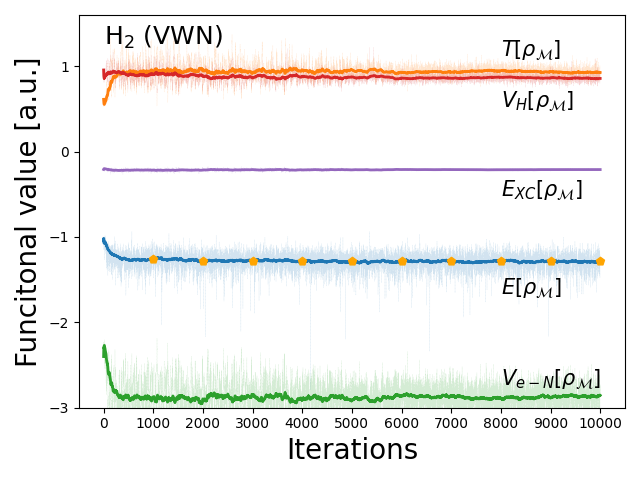}
    \includegraphics[scale=0.3]{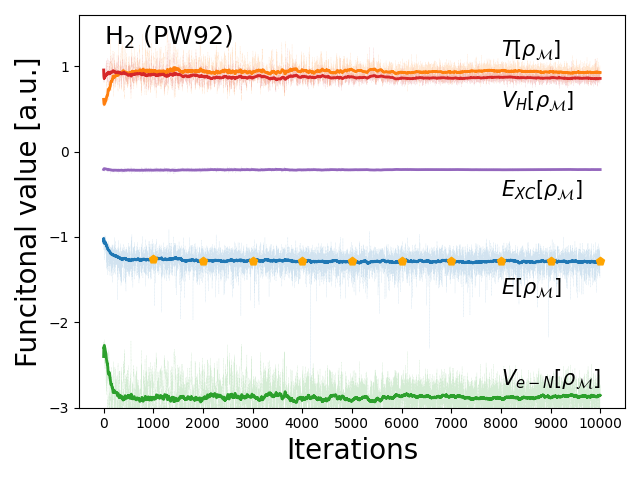}
    \includegraphics[scale=0.3]{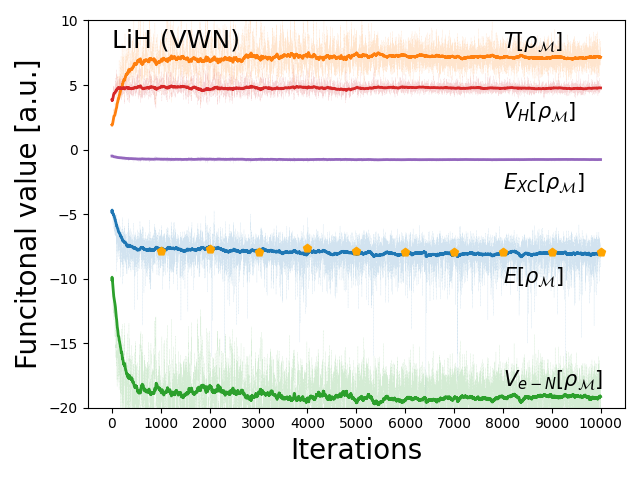}
    \includegraphics[scale=0.3]{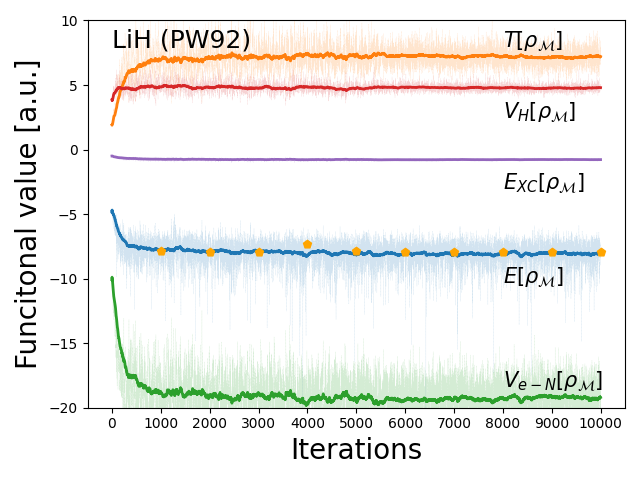}
    \includegraphics[scale=0.3]{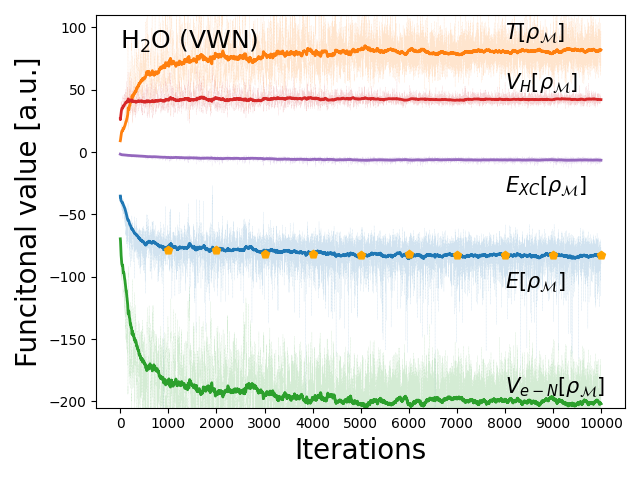}
    \includegraphics[scale=0.3]{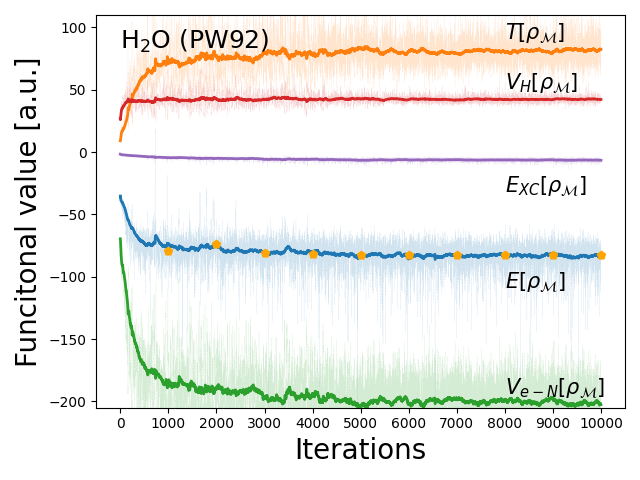}
    \caption{The value of $E[\rhom]$ and each functional, computed with MC (Eq. \ref{eqn:general_expectation}), at each iteration of the optimization. In this case, we considered a promolecular density ($\rhozeropm$) as our base distribution. The {\color{orange} $\pentagofill$}-symbols represent the value of the total energy computed with quadrature integration. For all simulations, we used the same architecture for $\gphi$ and the RMSProp optimizer, see the text for more details.}
    \label{fig:ema_mol}
\end{figure}

\begin{figure}[!ht]
    \centering
    \includegraphics[scale=0.3]{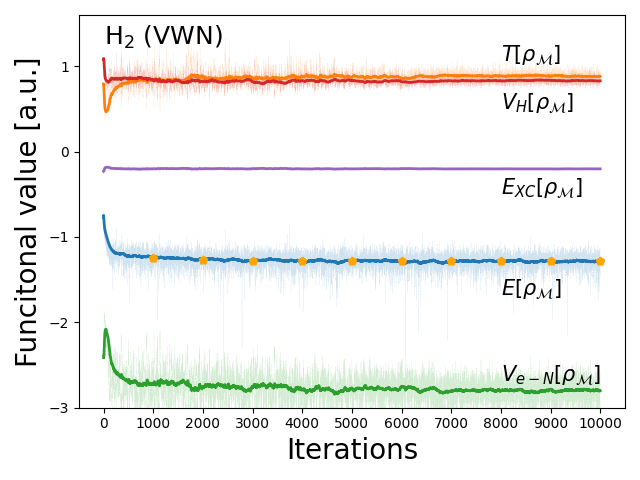}
    \includegraphics[scale=0.3]{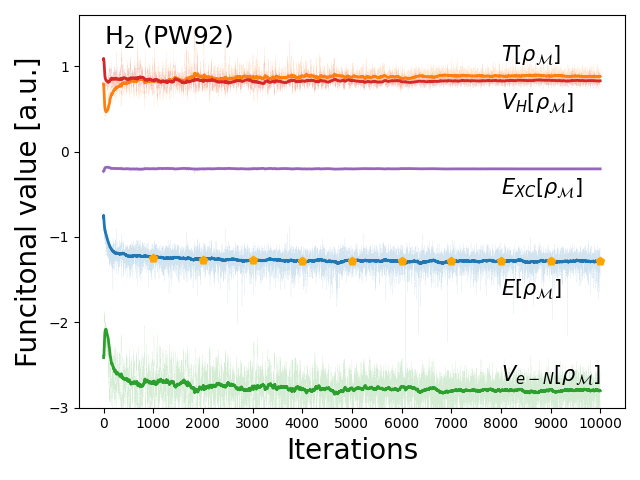}
    \includegraphics[scale=0.3]{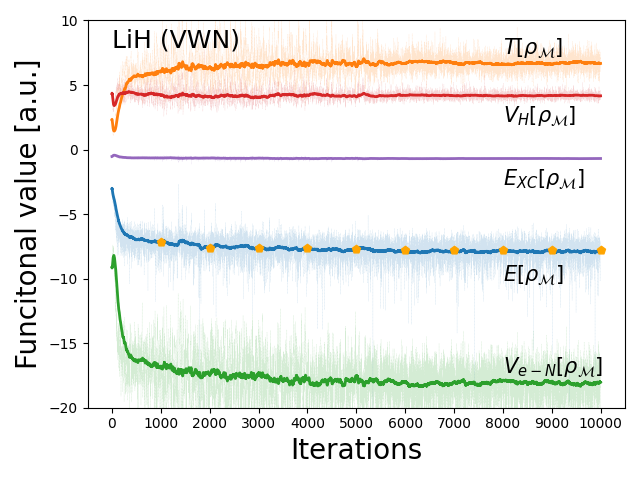}
    \includegraphics[scale=0.3]{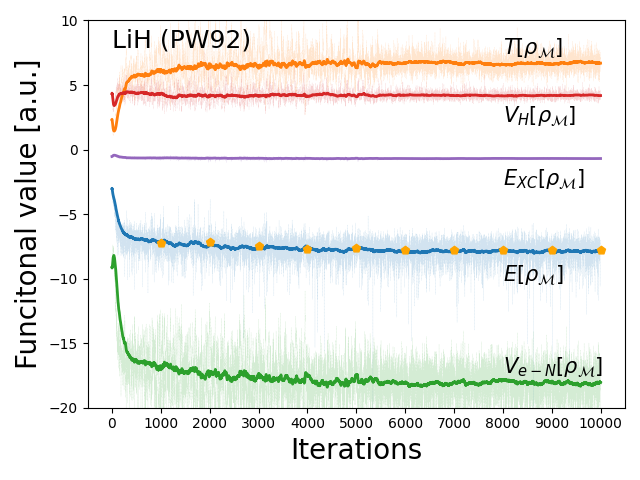}
    \includegraphics[scale=0.3]{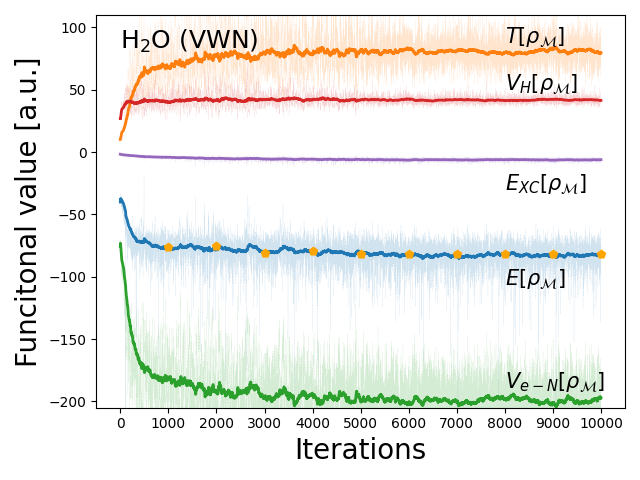}
    \includegraphics[scale=0.3]{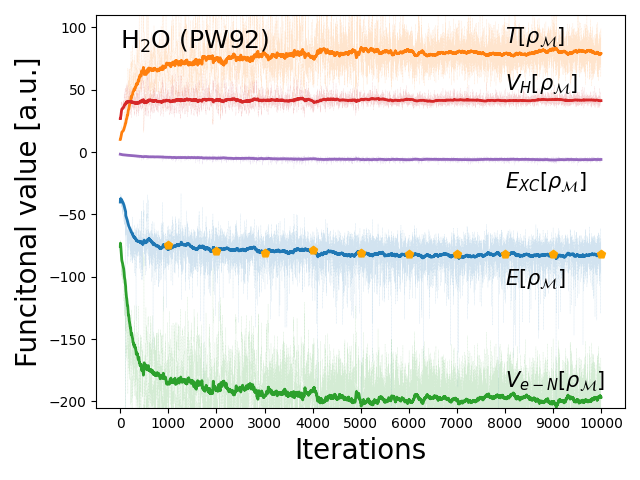}
    \caption{The value of $E[\rhom]$ and each functional, computed with MC (Eq.~\ref{eqn:general_expectation}), at each iteration of the optimization. In this case, we considered a single Gaussian distribution ($\rhozero$) as our base distribution. The {\color{orange} $\pentagofill$}-symbols represent the value of the total energy computed with quadrature integration. For all simulations, we used the same architecture for $\gphi$ and the RMSProp optimizer, see the text for more details.}
    \label{fig:ema_uni_mol}
\end{figure}

\begin{figure}[!ht]
    \centering
   
        \centering
        
        \includegraphics[scale=0.23]{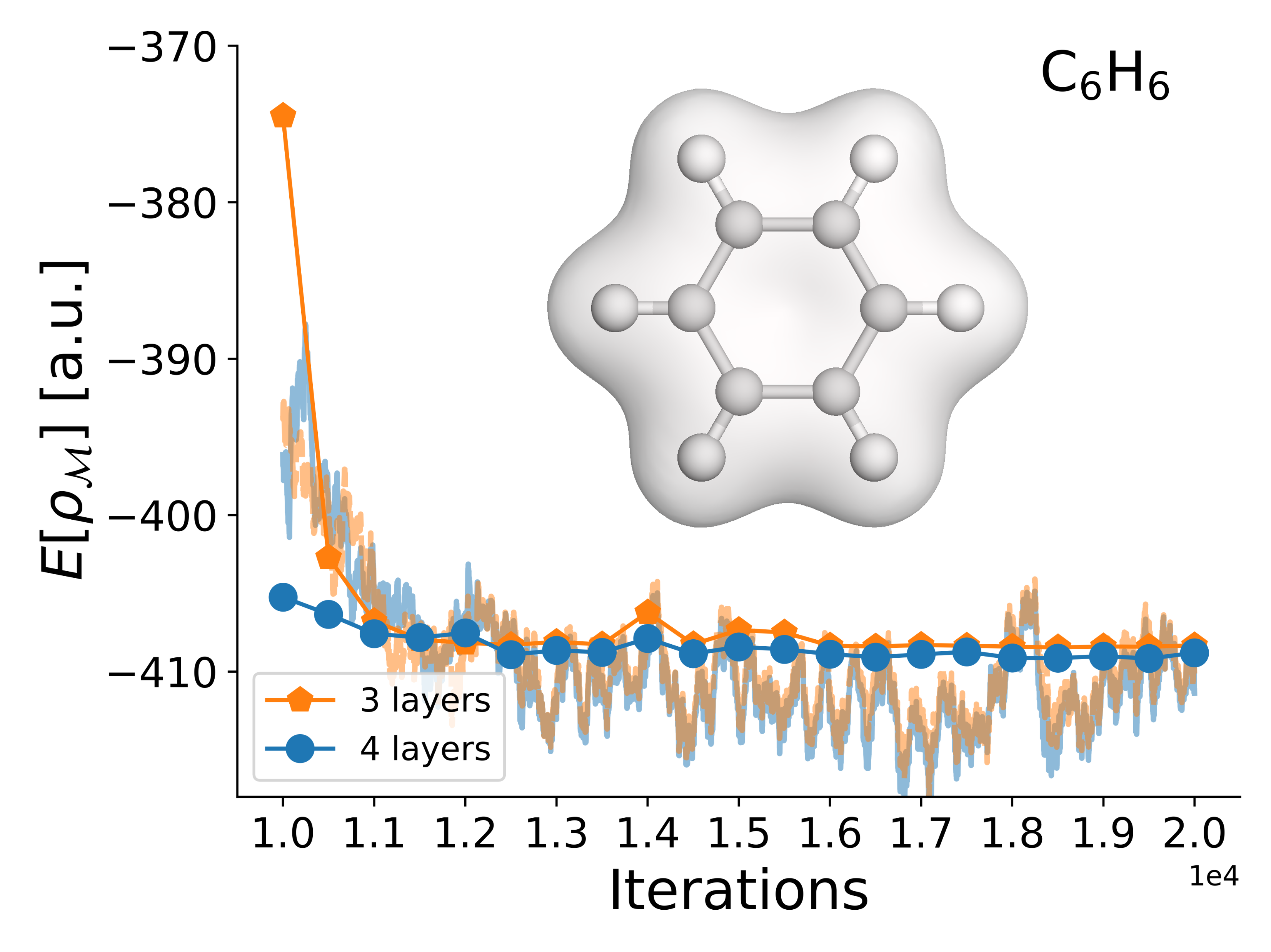}
        \includegraphics[scale=0.23]{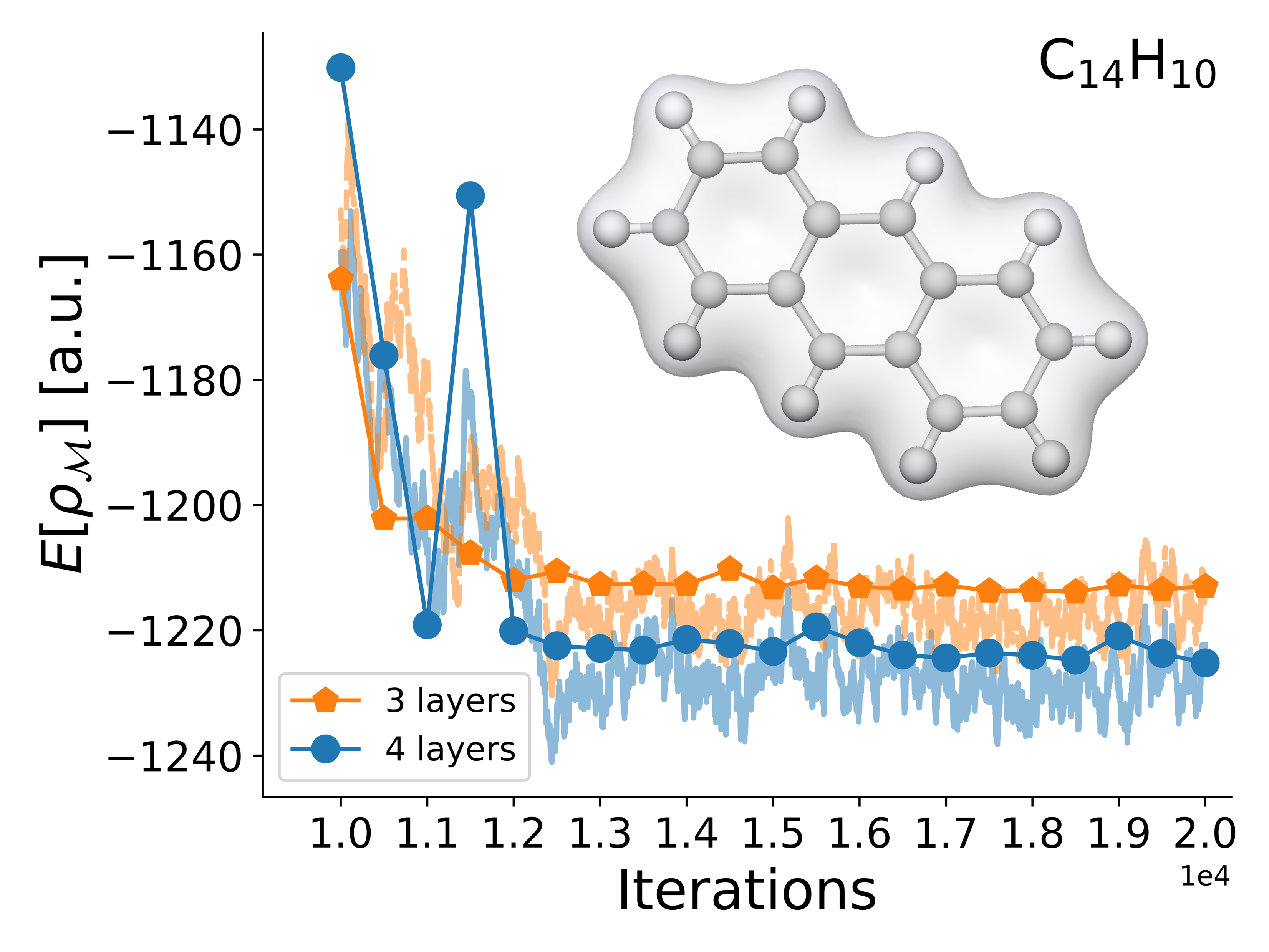}\\
        \includegraphics[scale=0.23]{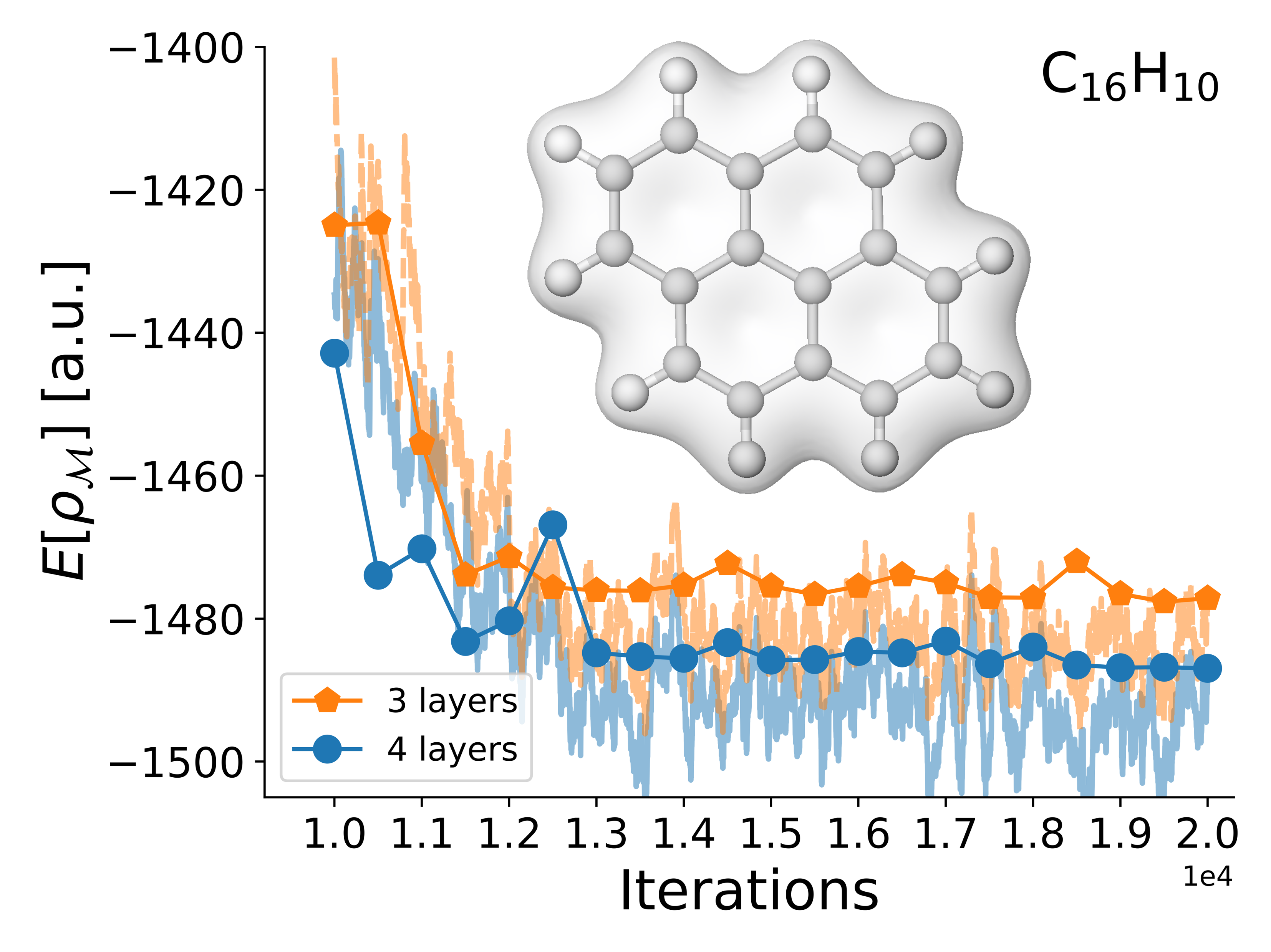}
        \includegraphics[scale=0.23]{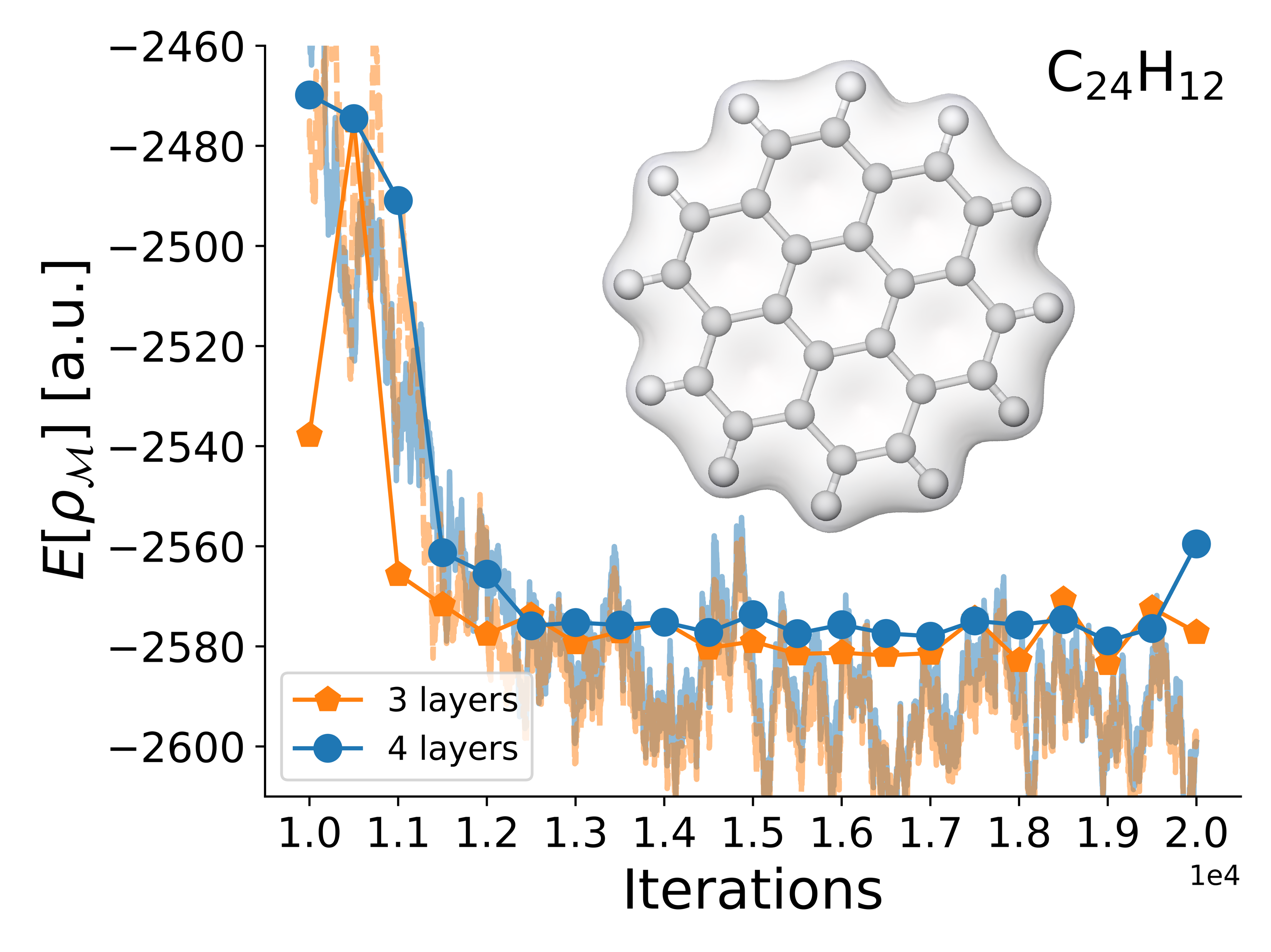}
      
        \caption{The total energy of the molecule through the optimization for a CNF with a promolecular density ($\rhozeropm$). The symbols indicate the total energy computed with quadrature integration and the curves with Monte Carlo. For these simulations, we considered a GNN (Eq. \ref{eqn:equiv_nn}) with 3 ({\color{orange} $\pentagofill$}) and 4 layers ({\color{blue} $\bullet$}) with 64 neurons per layer, the PW92 functional and Adam optimizer.
    We also display the electron density parametrized with a CNF with the lowest total energy computed with quadrature integration. }\label{fig:cxhx_figureds_ema}
  
\end{figure}

The total energy functional was composed of the sum of $T_{\text{TF}}[\rhom]$, $T_{\text{W}}[\rhom]$, $V_{H}[\rhom]$, and $V_{\text{e-N}}[\rhom]$ (Eqs.~\ref{eqn:exp_TF}-\ref{eqn:exp_e-N}). For the XC functional, we combine $\epsilonxlda$ and $\epsilonxbe$, Eqs.~\ref{eqn:x_Dirac}-\ref{eqn:x_b88} respectively, as the exchange component, and intercalate $\epsilon_{\text{C}}^{\text{VWN}}$ (Eq. ~\ref{eqn:correlation_vwn}) and $\epsilon_{\text{C}}^{\text{PW92}}$ (Eq.~\ref{eqn:correlation_pw92}) as the correlation term. Figs. S.\ref{fig:ema_mol}-S.\ref{fig:ema_uni_mol} show the value of each functional for H$_2$, LiH, and H$_2$O as a function of the iteration steps. For Fig. S.\ref{fig:ema_mol}, the base distribution was the promolecular density, and for Fig. S.\ref{fig:ema_uni_mol} $\rhozero$ was a single multi-variate Gaussian distribution. For both sets of simulations, $\gphi$'s architecture (Eq.~\ref{eqn:equiv_nn}) comprised three layers. 
Fig. S.\ref{fig:cxhx_figureds_ema} shows the value of the total energy ($E[\rhom]$) as a function of the iteration steps for the hydrocarbon molecules (C$_{x}$H$_{y}$ molecules) using the promolecular density. For this set of simulations, $\gphi$'s architecture (Eq.~\ref{eqn:equiv_nn}) comprised four layers. 

Table S.\ref{tab:3d_energies} contains the total energy values of all molecular systems considered here for the VWN and PW92 functionals using the promolecular density.
For all systems, we considered multiple layers to examine the impact of having a $\gphi$ with more layers. We also report the average iteration step time using a minibatch size of 512 for the small molecules and 724 for the hydrocarbon molecules (C$_{x}$H$_{y}$ molecules).  
For bigger systems, we found that the training procedure reaches an equilibrium after $\sim 12,000$ training steps, as observed in Fig. S.\ref{fig:cxhx_figureds_ema}.
To validate the proposed methodology, we computed the total energy employing quadrature integration methods used in DFT packages~\citep{becke1988multicenter}. These results are also presented in Table S.\ref{tab:3d_energies}.

\begin{figure}[!ht]
    \centering
        \centering
    \includegraphics[scale=0.3]{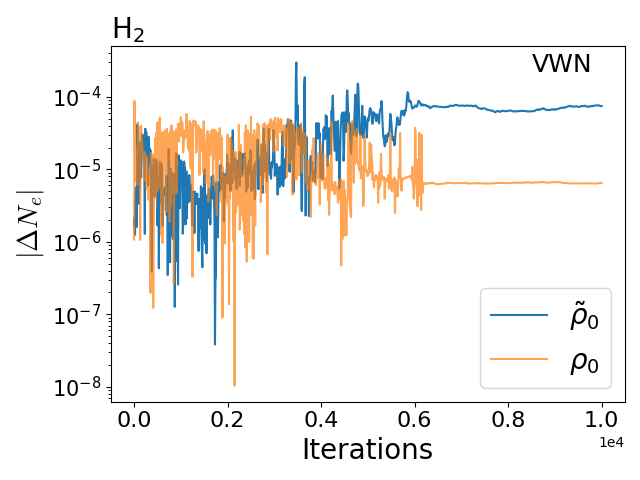}
    \includegraphics[scale=0.3]{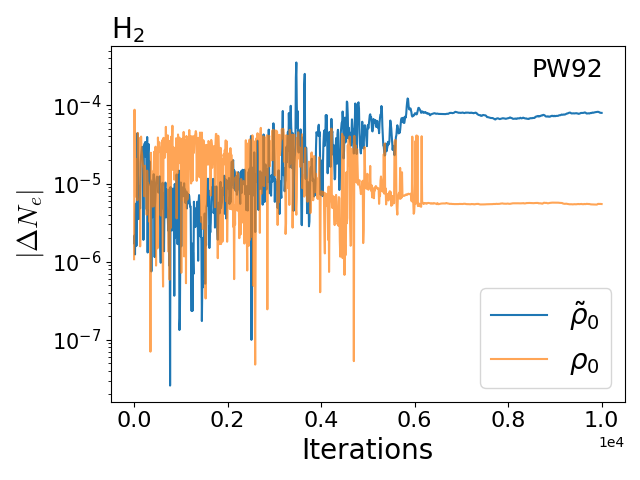}
    \includegraphics[scale=0.3]{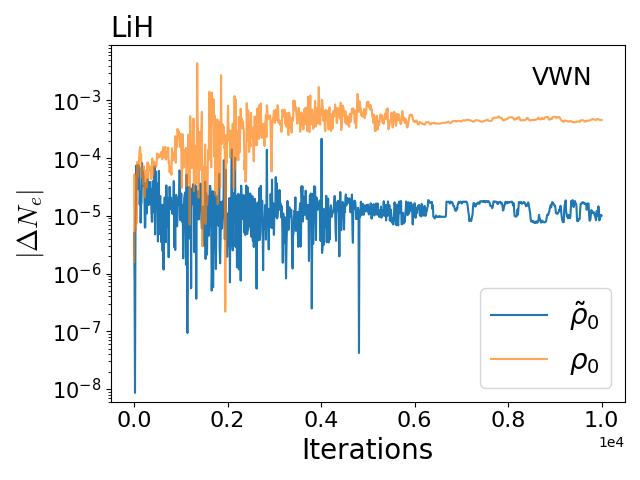}
    \includegraphics[scale=0.3]{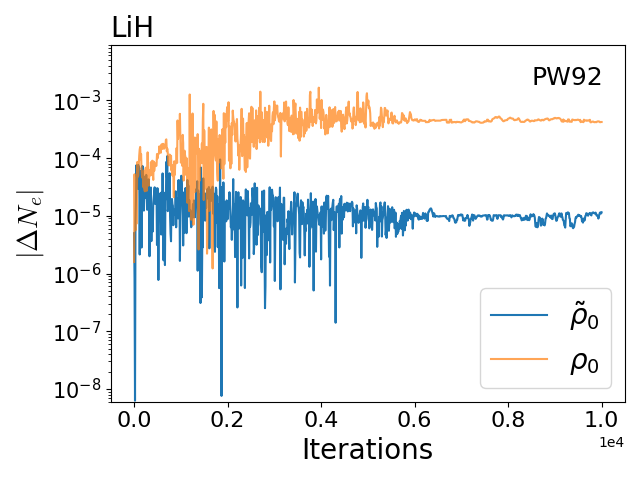}
    \includegraphics[scale=0.3]{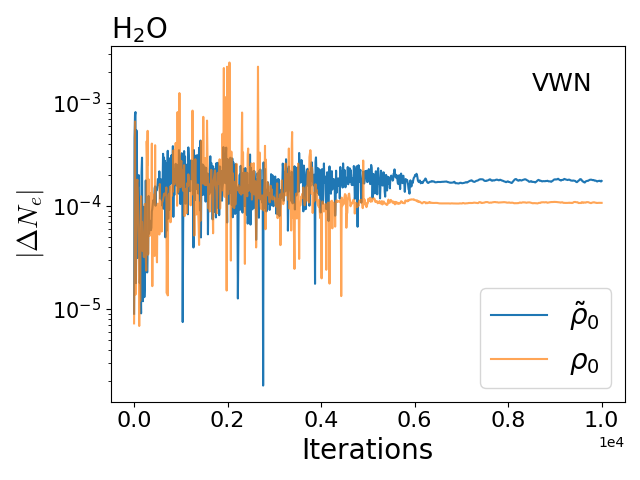}
    \includegraphics[scale=0.3]{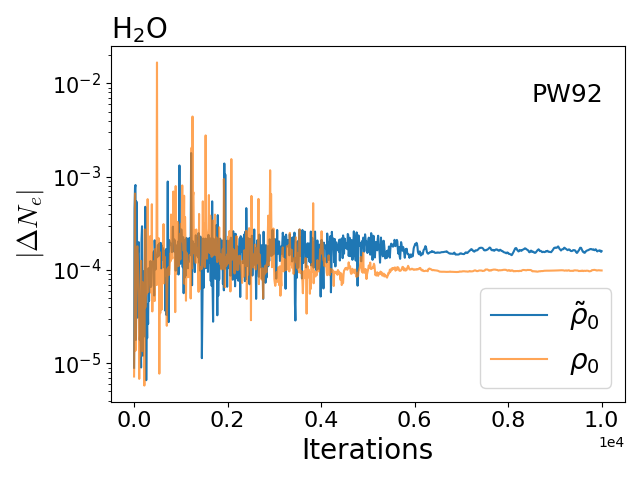}
    
    \caption{The absolute difference between the number of electrons ($\Ne$) and the integral of $\rhom$ at each iteration during the optimization.}
    \label{fig:norm_mol}
   
\end{figure}

To verify the normalization of $\rhom$, we also used quadrature integration methods to compute the difference with respect to the $\Ne$,
\begin{equation}
    \Delta\Ne = \int \rhom(\vx) \mathrm{d}\vx - \Ne.
    \label{eqn:normalization_guarantee}
\end{equation}
Fig. S.\ref{fig:norm_mol} shows the value of $\Delta\Ne$ for the small molecules and Table S.\ref{tab:delta_Ne} reported the absolute value of $\Delta\Ne$ for the C$_{x}$H$_{y}$ molecules.

\begin{table}[]
    \centering
    \begin{tabular}{|c|c|c|c|c|c|}
    \hhline{------}
    \multicolumn{2}{|c|}{}&\multicolumn{2}{c|}{Adam}&\multicolumn{2}{c|}{RMSProp}\\
    \hline
    \multirow{1}{*}{Molecule ($\Ne$)}& \# Layers&VWN&PW92&VWN&PW92\\
    \hline
    \multirow{2}{8em}{ \centering C$_{6}$H$_{6}$ ($42$) benzene} 
    &$3$&$5.0\times 10^{-5}$&$7.5\times 10^{-5}$&$5.3\times 10^{-5}$&$9.4\times10^{-5}$ \\
    &$4$&$2.7\times10^{-5}$&$1.1\times10^{-5}$&$3.8\times10^{-5}$&$3.5\times10^{-5}$\\
    \hline
    \multirow{2}{8em}{ \centering C$_{14}$H$_{10}$ ($96$) anthracene}&$3$&$6.4\times10{-5}$&$5.5\times10^{-5}$&$1.8\times10{-4}$&$1.5\times10^{-4}$\\ &$4$&$1.4\times10^{-4}$&$1.9\times10^{-4}$&$1.1\times10^{-4}$&$1.7\times10^{-4}$\\ 
    \hline
    \multirow{2}{8em}{ \centering C$_{16}$H$_{10}$ ($106$) pyrene}
    &$3$&$1.0\times10^{-4}$&$8.7\times10^{-5}$&$2.7\times10^{-4}$&$2.9\times10^{-4}$ \\
    &$4$&$2.4\times10^{-4}$&$2.1\times10^{-4}$&$4.9\times10^{-4}$&$5.1\times10^{-4}$ \\ 
    \hline
    \multirow{2}{8em}{\centering C$_{24}$H$_{12}$ ($156$) coronene}
    &$3$&$6.9\times10^{-4}$&$6.5\times10^{-4}$&$ 6.3\times10^{-4}$&$6.0\times10^{-4}$ \\ 
    &$4$&$7.2\times10^{-4}$&$9.5\times10^{-4}$&$7.9\times10^{-4}$&$8.4\times10^{-4}$ \\
    \hline
    
    \end{tabular}
    \caption{The absolute value of $\Delta\Ne$ (Eq. \ref{eqn:normalization_guarantee}) for the hydrocarbon molecules (C$_{x}$H$_{y}$ molecules).}
    \label{tab:delta_Ne}
\end{table}

\begin{sidewaystable}

\scriptsize
\begin{longtable}{|c|c|c|c|c|c|c|c|c|c|c|c|c|}
    \hhline{------}
     
    \hhline{|------------}
    \multicolumn{2}{|c}{}&\multicolumn{4}{c|}{Total energy MC integration}&\multicolumn{4}{c|}{Total energy quadrature integration}&\multicolumn{2}{|c|}{Iteration time [s]}\\
    \hhline{|------------}
    \multicolumn{2}{|c|}{}&\multicolumn{2}{c|}{Adam}&\multicolumn{2}{c|}{RMSProp}&\multicolumn{2}{c|}{Adam}&\multicolumn{2}{c|}{RMSProp}&\multicolumn{2}{c|}{Adam/RMSProp}\\
    
    \hhline{|------------|}
    \multirow{1}{*}{Molecule ($\Ne$)}& \# Layers&VWN&PW92&VWN&PW92&VWN&PW92&VWN&PW92&VWN&PW92\\
    \hline
    \multirow{3}{*}{H$_2$ ($2$)}
    &$1$&$-1.28043$&$-1.28040$&$-1.28038$&$-1.28045$&$-1.27668$&$-1.27665$&$-1.27666$&$-1.27674$&$0.12(2)$&$0.12(2)$\\
    &$2$&$-1.28860$&$-1.28854$ &$-1.28811$&$-1.28823$&$ -1.28501$&$-1.28504$&$-1.28463$&$-1.28475$&$0.28(5)$&$0.29(5)$\\
    &$3$&$-1.29075$&$-1.29078$&$-1.29053$&$-1.29055$&$-1.28716 $&$-1.28723$&$-1.28703$&$-1.28704$&$0.44(7)$&$0.44(7)$\\
    \hline
    \multirow{3}{*}{LiH ($4$)}&$1$&$-7.95421$&$-7.95479$&$-7.94972$&$-7.95175$&$-7.84666$&$-7.84684$&$-7.84346$&$-7.84428$&$0.16(3)$&$0.16(3)$\\
    &$2$&$-8.06433$ &$-8.05864$ &$-8.05514$&$-8.06152$&$-7.95530$&$-7.95300$&$-7.95066$&$-7.95185$&$0.34(6)$&$0.35(5)$\\&$3$&$-8.07509$&$-8.07257$&$-8.06180$&$-8.06198$&$-7.96105$&$-7.96297$&$-7.96081$&$-7.96262$&$0.52(8)$&$0.53(7)$\\
    \hline
    \multirow{3}{*}{H$_2$O ($10$)}
    &$1$&$-79.76794$&$-79.70309$&$-79.58261$&$-79.57113$&$-78.59261$&$-78.57637$&$-78.40929$&$-78.37367$&$0.27(5)$&$0.29(6)$\\
    &$2$&$-82.77371$&$-82.91470$&$ -82.80063$&$-82.77893$&$-81.96175$&$-82.16772$&$-82.00596$&$-81.96604$&$0.56(8)$&$0.55(7)$\\
    
    &$3$&$-83.23742$&$ -83.38899$&$-83.27477$&$-83.25400$&$-82.44516$&$-82.48210$&$-82.41970$&$-82.38154$&$0.81(9)$&$0.86(9)$\\
    \hline
    \multirow{2}{8em}{ \centering C$_{6}$H$_{6}$ ($42$) benzene}
    &$3$&$-409.84658$&$-409.63297$&$-410.44893$&$-410.17949$&$-408.46554$&$-408.49605$&$ -409.26159$&$ -409.05600$&$4.5(7)$&$4.7(8)$\\
    &$4$&$-410.61044$&$-410.03415$&$ -413.74284$&$-413.60553$&$-409.47686$&$-409.19518$&$-412.29239$&$-412.17791$&$6(1)$&$6.7(9)$\\
    \hline
    \multirow{2}{8em}{ \centering C$_{14}$H$_{10}$ ($96$) anthracene}
    &$3$&$-1219.39095$&$ -1216.54545$&$-1201.12833$&$-1203.82600$&$-1216.8759$&$-1214.16513$&$-1198.22969$&$-1200.47439$&$8(2)$&$7(1)$\\
    &$4$&$-1227.42202$&$-1227.93900$&$-1217.00866$&$-1215.75897$&$-1224.91707$&$-1225.25808$&$-1213.72391$&$-1212.40638$&$9(2)$&$9(2)$\\
    \hline
    \multirow{2}{8em}{ \centering C$_{16}$H$_{10}$ ($106$) pyrene}
    &$3$&$-1485.36895$&$ -1481.67124$&$-1462.95408$&$-1460.88932$&$-1480.2493$&$-1477.8062$&$-1459.9878$&$-1457.1779$&$9(2)$&$8(2)$\\
    &$4$&$-1488.20147$&$-1491.04825$&$-1481.22059$&$-1481.20265$&$-1484.54730$&$-1487.03938$&$-1476.59673$&$-1475.88246$&$10(2)$&$10(2)$\\
    \hline
    \multirow{2}{8em}{\centering C$_{24}$H$_{12}$ ($156$) coronene}
    &$3$&$-2601.68518$&$ -2604.28162$&$-2541.68127$&$-2545.02514$&$-2581.70553$&$-2583.91283$&$-2519.84625$&$-2523.29440$&$13(3)$&$11(3)$\\
    &$4$&$-2601.88376$&$-2602.41654$&$-2560.61346$&$-2560.74780$&$-2580.25980$&$ -2580.03025$&$-2538.83560$&$-2537.96701$&$17(3)$&$15(3)$\\
    \hline
    \caption{The total energy, reported in Ha, computed with MC and quadrature integration for two different XC functionals.  For the MC values, we averaged the lowest $300$ energy points. We also reported the average iteration-step time in seconds. } 
\label{tab:3d_energies}\\
\end{longtable}

\end{sidewaystable}

\newpage
\bibliography{library}